\documentclass[twocolumn,twocolappendix]{aastex7}
\usepackage{graphicx}
\usepackage{color}

\def\ltsima{$\; \buildrel < \over \sim \;$}
\def\simlt{\lower.5ex\hbox{\ltsima}}
\def\gtsima{$\; \buildrel > \over \sim \;$}
\def\simgt{\lower.5ex\hbox{\gtsima}}
\def\kms{\mbox{km s$^{-1}$}}

\def\cm2{\mbox{$\mbox{cm}^{-2}$}}
\def\cm3{\mbox{$\mbox{cm}^{-3}$}}
\def\h2{\mbox{$_{\mbox{\tiny H2}}$}}
\shorttitle{High-Resolution Mid-Infrared Absorption Imaging of the NGC6334 Filament with JWST}
\shortauthors{Andr\'e et al.}
\graphicspath{{./}{aastex631/figures/}}

\begin{document}

\title{Structure and Fragmentation Scale of a Massive Star-Forming Filament in NGC6334:\\
High-Resolution Mid-Infrared Absorption Imaging with JWST}

\author[0000-0002-3413-2293]{Philippe Andr\'e}
\affiliation{Laboratoire d'Astrophysique (AIM) \\ 
Universit\'e Paris-Saclay, Universit\'e Paris Cit\'e, CEA, CNRS, AIM\\
91191 Gif-sur-Yvette, France}
\email[show]{pandre@cea.fr}  

\author[0009-0005-9197-6483]{Michael Mattern}
\affiliation{Laboratoire d'Astrophysique (AIM) \\ 
Universit\'e Paris-Saclay, Universit\'e Paris Cit\'e, CEA, CNRS, AIM\\
91191 Gif-sur-Yvette, France}
\email{michael.mattern@cea.fr}

\author{Doris Arzoumanian}
\affiliation{National Astronomical Observatory of Japan \\
Osawa 2-21-1 \\ 
Mitaka, Tokyo 181-8588, Japan}
\email{doris.arzoumanian@nao.ac.jp}

\author[0000-0001-9368-3143]{Yoshito Shimajiri}
\affiliation{Kyushu Kyoritsu University\\ 
Jiyugaoka 1-8, Yahatanishi-ku\\ 
Kitakyushu, Fukuoka, 807-08585, Japan}
\email{yoshito.shimajiri@gmail.com}

\author[0000-0001-9509-7316]{Annie Zavagno}
\affiliation{Aix Marseille Univ, CNRS, CNES, LAM \\ 
Marseille, France}
\email{annie.zavagno@lam.fr}

\author[0000-0001-6891-2995]{Daisei Abe}
\affiliation{Astronomical Institute, Tohoku University\\ 
Sendai, Miyagi 980-8578, Japan}
\email{abe.daisei@astr.tohoku.ac.jp}

\author{Delphine Russeil}
\affiliation{Aix Marseille Univ, CNRS, CNES, LAM \\ 
Marseille,France}
\email{delphine.russeil@lam.fr}



\begin{abstract}

{Dense filaments are believed to be representative of the initial conditions of star formation in molecular clouds. We have used the MIRI instrument on JWST to image the massive 
filament NGC~6334M at $d \sim 1.3$\,kpc with unprecedented resolution and dynamic range at 7.7 and 25.5\,$\mu$m. Our observations reveal the fine structure of the filament in absorption 
against mid-infrared background emission. From the absorption data, we derive high-resolution column density maps and perform a detailed analysis of the structure of NGC~6334M. 
We find a median filament width of $0.12 \pm 0.02$\,pc at both wavelengths, resolved by {almost} two orders of magnitude with MIRI, and consistent with the typical half-power width 
of {\it Herschel} filaments in nearby  ($d < 500$\,pc) clouds. The JWST data also reveal the presence of a quasi-periodic series of side filaments with a similar {projected} spacing of $0.125 \pm 0.015$\,pc. 
Combining our JWST results with {\it Spitzer} and APEX/{\it Herschel} data, we perform a study of cloud structure over four orders of magnitude in linear scale. A convergence test shows that our width estimates for NGC~6334M are robust and reflect the presence of a true characteristic scale. While there is evidence of a Kolmogorov-like spectrum of small-scale fluctuations 
down the $1.6 \times 10^{-3}$\,pc resolution of the JWST observations, we identify a break in the power spectrum of column density fluctuations at a scale $\sim$0.1--0.4 pc comparable 
to the width of NGC~6334M and its side filaments. This characteristic scale $\sim$0.1\,pc has important implications for the origin of the star formation efficiency in dense gas and the IMF.
}

\end{abstract}

\keywords{stars: formation -- ISM: clouds -- ISM: structure  -- submillimeter: ISM}    


\section{Introduction} \label{sec:intro}

The filamentary structure of the cold interstellar medium (ISM) plays a central role in the star formation 
process \citep[e.g.,][for recent reviews]{Hacar+2023,Pineda+2023}. 
{\it Herschel} imaging surveys of nearby Galactic clouds \citep{Andre+2010,Molinari+2010,Hill+2011,Juvela+2012} 
have shown that filaments dominate the mass budget of molecular clouds at high densities \citep[e.g.,][]{Schisano+2014} 
and correspond to the birthsites of most prestellar cores 
\citep[e.g.,][]{Konyves+2015,Marsh+2016,DiFrancesco+2020}.  
The {\it Herschel} results suggest that molecular filaments are representative of the initial conditions of most 
star formation in the Galaxy \citep[e.g.,][]{Andre+2014}. 
Characterizing their physical properties is therefore paramount. 
An essential attribute of filamentary structures is their transverse diameter since the fragmentation properties of cylindrical filaments 
are expected to scale with the filament diameter, at least according to quasi-static fragmentation models 
\citep[e.g.,][]{Nagasawa1987,Inutsuka+1992}. 
Remarkably, observations of nearby molecular filaments as part of the {\it Herschel} Gould Belt survey (HGBS -- \citealp{Andre+2010}) 
point to a common filament width of $\sim$0.1\,pc, with small dispersion around this 
typical value  \citep[][]{Arzoumanian+2011,Arzoumanian+2019, KochRosolowsky2015}. 
In particular, from a detailed study of radial column density profiles in 8 nearby clouds at $d < 500\,$pc,  
\citet{Arzoumanian+2019} concluded  that all 599 filaments of their {\it Herschel} sample share approximately the 
same half-power width $\sim$0.1\,pc with a spread of a factor of $\sim$2.
This is the case independently of the mass per unit length $M_{\rm line}$, 
whether the filaments are {\it subcritical} with $M_{\rm line}   \lesssim 0.5\, M_{\rm line, crit}$,  
{\it transcritical} with  $0.5\, M_{\rm line, crit} \la M_{\rm line} \la 2\, M_{\rm line, crit}$, 
or  {\it thermally supercritical} with $M_{\rm line}  \ga 2\, M_{\rm line, crit}$, 
where $M_{\rm line, crit} = 2\, c_s^2/G $ is the thermal value of the critical mass per unit length \citep[e.g.,][]{Ostriker1964}, 
i.e., $\sim$$\, 16\, M_\odot \, {\rm pc}^{-1} $ 
for a sound speed $c_{\rm s} \sim 0.2$\,\kms 
or a gas temperature \hbox{$T \approx 10\,$K}. 
There has been some debate about the reliability 
of this result \citep{Panopoulou+2017, Panopoulou+2022}, 
but tests performed on synthetic data \citep{Arzoumanian+2019, Roy+2019, Andre+2022} 
suggest that {\it Herschel} width measurements are quite robust, 
at least for nearby ($d < 0.5$\, kpc), high-contrast filaments. 
If confirmed, the existence of a typical filament width would have far-reaching implications 
as it would introduce a characteristic length and therefore a mass scale in the structure of molecular clouds,  
which may have a direct bearing on the origin of the broad peak in the stellar initial mass function (IMF) \citep[cf.][]{Larson1985,Andre+2019}.  
An alternative view, however, is that the filamentary structure of molecular clouds is highly hierarchical in nature 
and essentially scale-free down to scales $\ll 0.1$\,pc \citep[e.g.,][]{Hacar+2023}.

Studying the structure of highly supercritical filaments (with $M_{\rm line} \gg M_{\rm line, crit}$) and their environment 
is especially important in this context as the mere existence of these filaments poses a challenge to theoretical models. 
While cylinder fragmentation theory predicts that isothermal filaments with $M_{\rm line}$ within a factor of $\sim$\, 2 of $M_{\rm line, crit}$ 
should fragment into cores with a typical spacing of $\sim \,$4 times the filament width, 
highly supercritical filaments are expected to collapse radially to spindles in only about one free-fall 
time (or $< 10^5\, $yr) without significant fragmentation along their axis \citep{Inutsuka+1992, Inutsuka+1997}.
In contrast, {\it Herschel} observations reveal the presence of numerous $\sim$\,0.1-pc-wide supercritical filaments 
with widespread fragmentation into prestellar cores and an estimated lifetime $\sim 1\,$Myr 
\citep[e.g.,][]{Arzoumanian+2019,Konyves+2015,Andre+2014}. 
Since at least in the solar neighborhood most stars appear to form in dense $\sim$\,0.1-pc-wide filaments \citep[e.g.,][]{Konyves+2015}, 
investigating how thermally supercritical filaments may maintain a typical inner width $\sim 0.1$\,pc and fragment into prestellar cores 
instead of collapsing radially to spindles is crucial to understanding star formation. 
Although a promising theoretical model (``STORM'') has recently been proposed to account for such 
supercritical filaments based on non-ideal MHD simulations \citep{Abe+2025}, 
much work remains to be done before a full understanding of this problem is reached. 

An extreme example of a supercritical filament is the main filament in the high-mass star-forming  region NGC~6334, 
whose line mass approaches $M_{\rm line} \sim 1000\, M_\odot $/pc in its northern part \citep[cf. ][]{Andre+2016}.
At a distance of $\sim\, $1.3~kpc \citep{Chibueze+2014}, 
NGC~6334 is a very active complex of massive star formation \citep[][]{PersiTapia2008,Russeil+2013}
with about 150 associated luminous stars of O- to B3-type \citep[][]{Neckel1978,Bica+2003,Feigelson+2009}. 
At far-infrared and (sub)millimeter wavelengths, the central part of NGC\,6334 consists of a 10\,pc long elongated
structure including two major high-mass star-forming clumps and a narrow filament \citep[e.g.][]{Sandell2000, Andre+2016, Tige+2017}. 
The filament, which we will refer to as NGC~6334M in the following,  
is particularly prominent in ground-based (sub)millimeter continuum images 
where extended emission is effectively filtered out \citep[e.g.][]{Munoz+2007, Matthews+2008}. 
It apparently forms only low- to intermediate-mass stars \citep{Tige+2017}, except perhaps at its end points, 
in marked contrast with the high-mass clumps of the region 
which host several protostellar ``massive dense cores'' 
\citep{Sandell2000,Tige+2017}.
Following up on a 350\,$\mu$m continuum mapping of NGC~6334 
at $8\arcsec $ resolution with the ArT\'eMiS camera on APEX \citep{Andre+2016}, 
ALMA 3\,mm observations identified 26 compact  ($\sim\,$0.03\,pc) dense cores in the continuum 
and five velocity-coherent ``fiber-like'' sub-structures in N$_2$H$^+$ within the filament \citep{Shimajiri+2019b}.
The ALMA cores have a mass distribution peaking at $\sim 10\, M_\odot $, 
which is an order of magnitude higher than the peak of the prestellar core mass function (CMF) 
in nearby, low-mass star-forming clouds \citep{Konyves+2015,Marsh+2016}.

In this letter, we present the results of high-resolution (0.26\arcsec--0.84\arcsec) 7.7\,$\mu$m and 25.5\,$\mu$m imaging observations 
of the central part of NGC~6334M with the MIRI instrument on board JWST, which reveal the fine structure of the filament 
in absorption against bright mid-infrared background emission. 
The subarcsec resolution of JWST/MIRI, corresponding to 1.6--5.3$\times 10^{-3}$~pc at the distance of NGC~6334, 
allow us to resolve, for the first time, the transverse size of the main filament by nearly two orders of magnitude.  
In Sect.~\ref{sec:obs}, we describe our JWST/MIRI observations and use the resulting mid-infrared absorption data 
to construct high-resolution column density maps of NGC~6334M and its immediate environment.
In Sect.~\ref{sec:rad_prof}, we characterize the density structure of the filament by 
analyzing the distribution of radial column density profiles and deriving the autocorrelation function and 
transverse correlation length of the column density map(s).
In Sect.~\ref{sec:conv_test}, 
we perform a convergence test for the measured widths of NGC~6334M, similar to that presented 
by \citet{Panopoulou+2022} and \citet{Andre+2022} for the B211-3 filament in Taurus. 
In Sect.~\ref{sec:concl}, we compare our JWST findings on small $\la 1$\,pc scales to results obtained with {\it Herschel}
on larger scales up to $\sim$\,20\,pc, identify a break in the power spectrum of column density fluctuations at a scale
similar to the $\sim$\,0.12\,pc width derived for the filament, and conclude the paper.


\section{JWST Observations}
\label{sec:obs}

\begin{figure}
\centering
\resizebox{0.99\hsize}{!}{\includegraphics[angle=0]{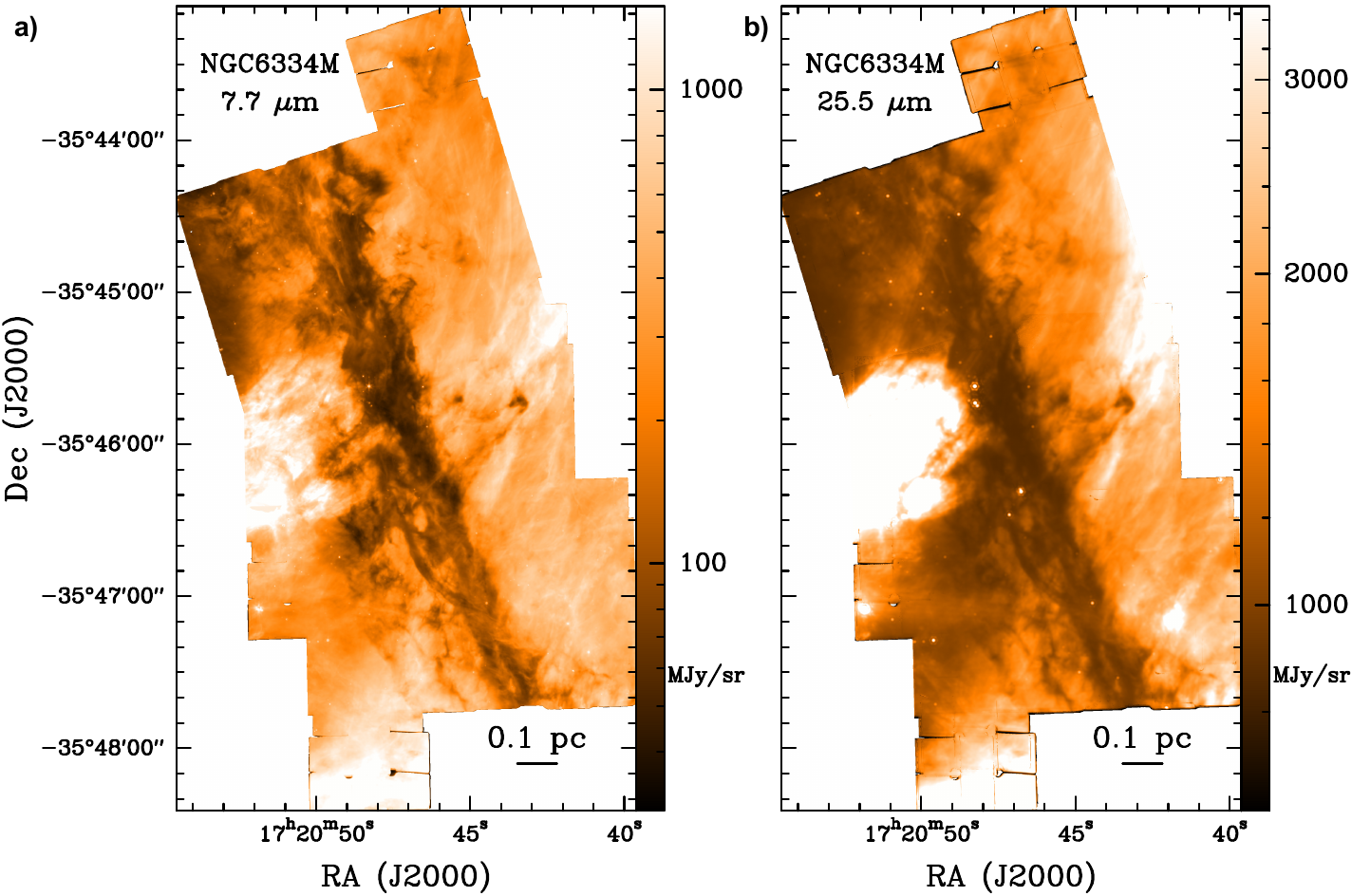}} 
\caption{JWST-MIRI images of the infrared-dark NGC6334M filament at 7.7\,$\mu$m {\bf (a)} and 25.5\,$\mu$m {\bf (b)}.
}
\label{ngc6334_jwst_images}
\end{figure}

The NGC~6334 filament was imaged under JWST Cycle 1 program 2526 (PI: Ph. Andr\'e; co-PI: M. Mattern) 
with the MIRI instrument in the F770W and F2550W broadband filters.  
{The data presented in this article were obtained from the Mikulski Archive for Space Telescopes (MAST) 
at the Space Telescope Science Institute. The specific JWST observations analyzed can be accessed 
via \dataset[doi: 10.17909/2ra1-ds91]{https://doi.org/10.17909/2ra1-ds91}.} 
They consist in a mosaic of three MIRI pointings with four dithered exposures in each of the two filters,   
were taken on 2023 August 14--15 and 2024 April 15, for a total of 7.3~hr. 
They were reduced using the publicly available reduction pipeline 
along with a few, additional in-house reduction steps described below. 
To combine all exposures into one mosaic, stage 3 of the reduction pipeline (version 1.14.0) was used. 
Due to the very low number of point sources (typically stars) in the observed region, 
the coordinate alignment of the pipeline is slightly mismatched by $\sim 0.05\arcsec$. 
This caused the default pipeline to flag 
pixels along strong intensity gradients in the overlapping region between the 2024 April and 2023 August exposures. 
The mean intensity of the available exposures at these positions was assigned to flagged pixels. 
The reduced mosaics cover a field of $\sim 3.6\arcmin \times 2.8\arcmin $ or $\sim1.4\,\rm{pc} \times 1.1\,{\rm pc} $ (see Figure~\ref{ngc6334_jwst_images}), 
have a typical rms instrumental noise of $ I_{\rm rms} \sim 0.3$--0.7\,MJy/sr (at 7.7--25.5\,$\mu$m)  
and achieve a half power beam width (HPBW) angular resolution 
of $\sim$\,$0.26\arcsec$ at 7.7\,$\mu$m and $\sim$\,$0.84\arcsec$ at 25.5\,$\mu$m.

\begin{figure}
\centering
\resizebox{0.99\hsize}{!}{\includegraphics[angle=0]{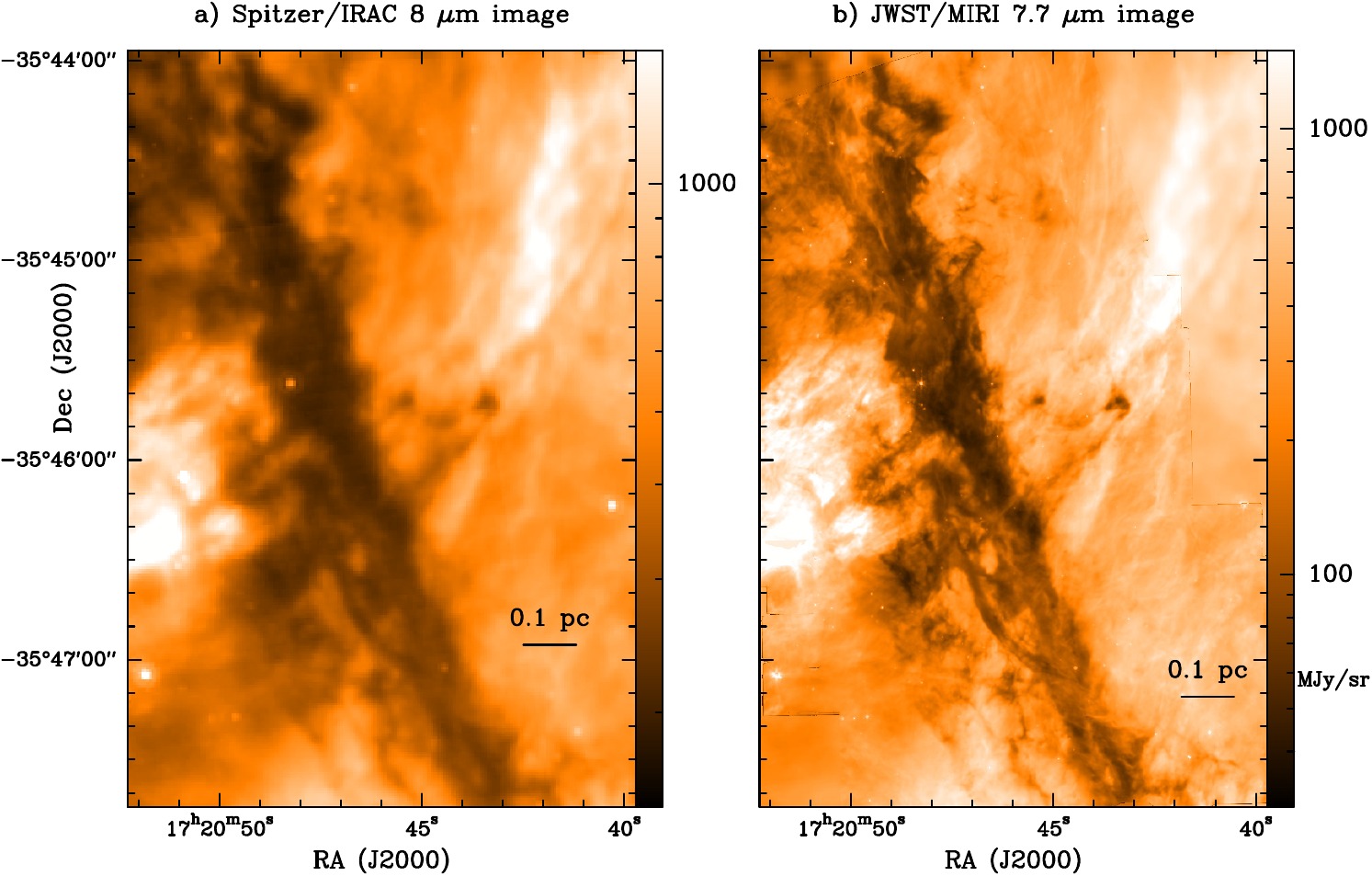}} 
\caption{Comparison between {\it Spitzer}-IRAC 8\,$\mu$m {\bf (a)} and JWST-MIRI 7.7\,$\mu$m {\bf (b)} images 
of the NGC6334M filament, with full width at half maximum (FWHM) resolutions of $\sim$$2\arcsec$ and $\sim$$0.26\arcsec$, respectively.
}
\label{ngc6334_spitzer_jwst_images}
\end{figure}

\begin{figure*}
\centering
\resizebox{0.98\hsize}{!}{\includegraphics[angle=0]{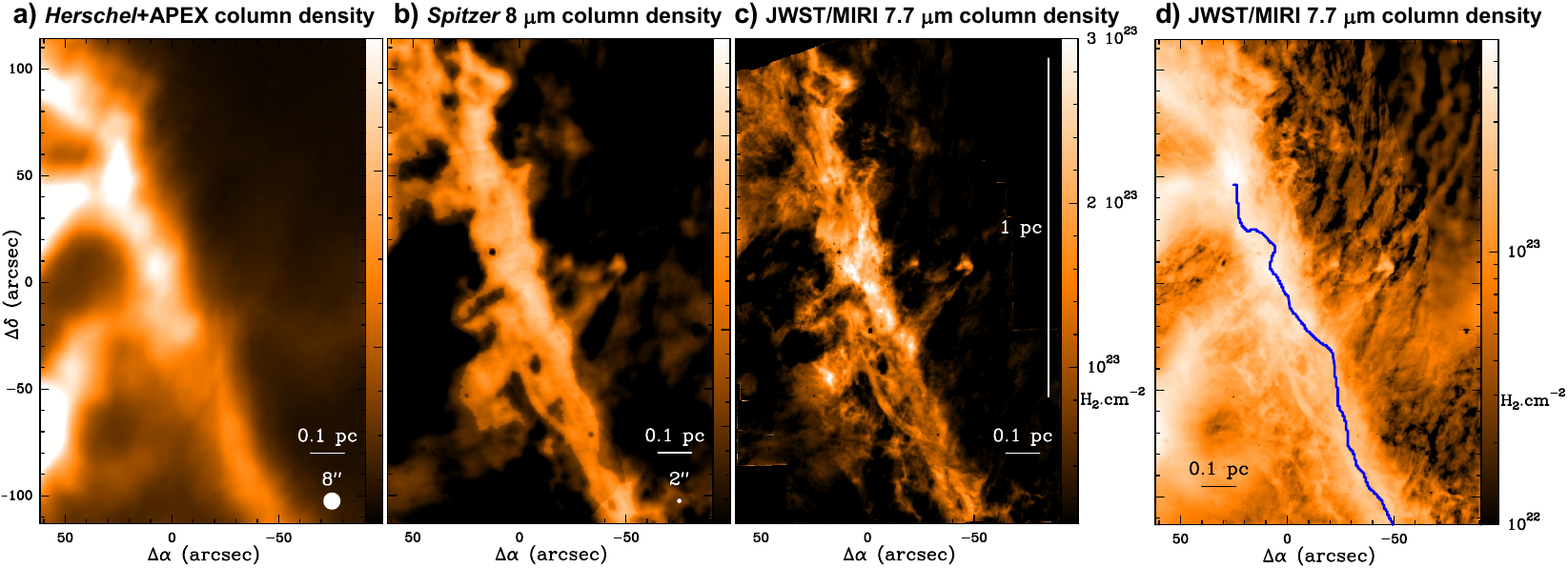}} 
\caption{Column density ($N_{H2}$) images of the infrared-dark NGC6334M filament 
derived from {\bf (a)} {\it Herschel}$+$APEX submm emission ($8\arcsec$ resolution); 
{\bf (b)} {\it Spitzer} 8\,$\mu$m absorption ($2\arcsec$ resolution) using method 2 in the text; 
{\bf (c)} JWST/MIRI\,7.7\,$\mu$m absorption ($0.26\arcsec$ resolution) using method 2 (map called $N_{\rm H_2}^{\rm JWST2}$ in the text); and 
{\bf (d)} JWST/MIRI\,7.7\,$\mu$m absorption ($0.26\arcsec$ resolution) using method 1 in the text ($N_{\rm H_2}^{\rm JWST1}$), 
with the filament crest derived by \citet{Andre+2016} from APEX/ArTéMiS 350\,$\mu$m data overlaid as a blue curve.
Panels  {\bf a)} to  {\bf c)} are displayed in linear scale and emphasize details at high column densities; 
panel {\bf d)} is displayed in logarithmic scale and emphasizes details at low column densities.
Note the {dramatic improvement in resolution and dynamic range with JWST} 
compared to APEX and {\it Spitzer}. 
}
\label{ngc6334_jwst_coldens_images}
\end{figure*}

\subsection{Comparison with earlier data and construction of high-dynamic-range column density maps}
\label{sec:coldens}

To illustrate the gain in angular resolution and image quality compared to {\it Spitzer} data, we show our MIRI 7.7\,$\mu$m image 
next to the IRAC 8\,$\mu$m view of the same region (from the GLIMPSE project -- \citealp{Churchwell+2009})
in Figure~\ref{ngc6334_spitzer_jwst_images}. 
When smoothed to the resolution of the {\it Spitzer}-IRAC data and slightly recalibrated to accommodate the slight difference in photometric bands, 
the JWST-MIRI  image is remarkably consistent with the  {\it Spitzer}-IRAC image. It is thus possible to combine both datasets and compensate for 
the limited field coverage of the JWST image, especially at large radial offsets from the spine of the filament.
Figure~\ref{ngc6334_spitzer_jwst_images}b shows a multiresolution 7.7\,$\mu$m image obtained from combining the original JWST image 
in Figure~\ref{ngc6334_jwst_images}a with the {\it Spitzer} image in Figure~\ref{ngc6334_spitzer_jwst_images}a. 
This multiresolution 7.7\,$\mu$m image has a resolution $\sim$\,$0.26\arcsec$ in the area covered by JWST and $\sim$\,$2\arcsec$ 
in the outer parts of the field only covered by {\it Spitzer}. Its dynamic range is quite remarkable, in terms of both surface brightness range 
and accessible spatial scales.

In both Figures~\ref{ngc6334_jwst_images} and ~\ref{ngc6334_spitzer_jwst_images}, the NGC~6334M filament shows up as a prominent dark lane.
The latter results from absorption of the bright mid-infrared background emission arising from both the Galactic Plane and the NGC~6334 complex itself
by the high column density of cold dust in the dense filament. 
More quantitatively, the observed emission $I_{\rm obs}$ at location $\vec{r}$ in the plane of sky can be expressed as: 
\begin{equation}
I_{\rm obs}(\vec{r})=I_{\rm back}(\vec{r}) \times e^{-\tau_\lambda}+I_{\rm fore}(\vec{r})
\label{equ1}
\end{equation}
where $I_{\rm back}$ is the mid-infrared intensity in the background of the absorbing structure, $I_{\rm fore}$ is the foreground intensity emitted in the line-of-sight 
between the filament and the observer, 
and $\tau_\lambda$ is the total optical depth produced by dust in the absorbing structure at wavelength $\lambda$ (here, either 7.7\,$\mu$m or 25\,$\mu$m). 
The optical depth $\tau_\lambda$ can then be derived as: 
\begin{equation}
\tau_\lambda={\rm ln} \frac{I_{\rm back}(\vec{r})}{I_{\rm obs}(\vec{r})-I_{\rm fore}(\vec{r})}.
\label{equ2}
\end{equation}
The column density, $N_{\rm H_2}$, of the absorbing structure is related to $\tau_\lambda$ by $ N_{\rm H_2}= \tau_\lambda / s_\lambda $,
where $s_\lambda$ is the dust extinction cross section. 
While the latter is not known to better than a factor of $\sim$\,2 at 7.7\,$\mu$m and 25\,$\mu$m, better accuracy in estimates 
of absolute column densities is not needed in the present study. 
Here, we adopt  \hbox{$s_{7.7\mu{\rm m}} = 2 \times 10^{-23}\, {\rm cm}^2$}, a value intermediate between the extinction cross sections used 
by \citet{Bacmann+2000} and \citet{Butler+2012} based on the dust models of \citet{Draine+1984} and \citet{Weingartner+2001}, respectively. 
Our adopted $s_{7.7\mu{\rm m}}$ value is consistent with the extinction law derived  by \cite{Rosenthal+2000} for the Orion OMC-1 cloud 
using observations of infrared H$_2$ lines with ISO (see Figure~4 of \citealp{Draine2003}). 
We also take \hbox{ $s_{25.5\mu{\rm m}} \approx s_{7.7\mu{\rm m}}$} based on the overall shape 
of the mid-infrared dust extinction curve \citep[e.g.,][]{Moneti+2001}.

We employed two methods for estimating the background and foreground intensities $I_{\rm back}$, $I_{\rm fore}$ 
and deriving high-resolution column density maps from the observed JWST images. 
The first method assumes that $I_{\rm fore}$ corresponds to the minimum value $I_{\rm obs}^{\rm min}$ of $I_{\rm obs}(\vec{r})$ in the mapped area 
(or in the vicinity of the absorbing structure) and relies on calibrating the distribution of background intensities $I_{\rm back}(\vec{r})$
by imposing that the derived absorbing column densities match the column densities obtained from independent submillimeter emission data 
at several positions 
{or over a full map} 
\citep[see][]{Bacmann+2000}.
Here, we used the $8\arcsec$-resolution column density map of the NGC~6634 region produced from 
APEX/ArT\'eMiS and {\it Herschel} submillimeter data 
as part of the CAFFEINE project \citep{Andre+2016,Mattern+2024} 
as our reference calibration map\footnote{The reference map adopted here is an improved version of the column density map 
derived by \citet{Andre+2016}, using additional ArT\'eMiS 350 and 450\,$\mu$m data taken at APEX in 2024 September 
in the context of the CAFFEINE project (\hbox{https://sites.google.com/view/artemis-apex-caffeine}). 
Combining these ArT\'eMiS data with {\it Herschel} observations at 160--500\,$\mu$m, a column density map at $8\arcsec$ resolution 
was constructed following the method described by \citet{Mattern+2024}, 
assuming the dust opacity law 
\hbox{$\kappa_{\lambda} = 0.1 \times (N_{\rm H_2}/N^0_{\rm H_2})^{0.28} \times (\lambda/300~\mu \rm m)^{-2}\ {\rm cm}^2 {\rm g}^{-1}$}
at submillimeter wavelengths with $N^0_{\rm H_2} = 1.7 \times 10^{22}\, {\rm cm}^{-2} $, in agreement with 
\citet[][see also \citealp{Schuller+2021}]{Roy+2013,Roy+2014}.}.
After smoothing the JWST image to the $8\arcsec$ resolution of the CAFFEINE map, we estimated 
the smoothed distribution of background intensities ${\tilde I_{\rm back}}(\vec{r})$ as:  
\begin{equation}
{\tilde I_{\rm back}}(\vec{r}) = \left(\tilde{I}_{\rm obs}(\vec{r}) - I_{\rm obs}^{\rm min} \right) \times  e^{N_{\rm H_2}^{\rm ArT}\, s_\lambda},
\label{equ3}
\end{equation}
where $\tilde{I}_{\rm obs}(\vec{r})$ and $N_{\rm H_2}^{\rm ArT}$ represent the smoothed JWST map and the CAFFEINE/ArT\'eMiS column density map, 
respectively. 
Following Equation~(\ref{equ2}), 
we then derived our first estimate of the absorbing column density distribution at JWST resolution $N_{\rm H_2}^{\rm JWST1}$ from: 
\begin{equation}
N_{\rm H_2}^{\rm JWST1}(\vec{r})\, s_\lambda = {\rm ln} \frac{{\tilde I_{\rm back}}(\vec{r})}{I_{\rm obs}(\vec{r})-I_{\rm obs}^{\rm min}}.
\label{equ4}
\end{equation}
This first approach assumes that the foreground emission is uniform in the vicinity of the absorbing structure and that the background emission 
has no significant fluctuations on angular scales smaller than the $8\arcsec$ resolution of the reference column density map. 
Its great advantage is that it produces a high-resolution column density image which is -- by design -- consistent with 
the lower-resolution reference column density map, here $N_{\rm H_2}^{\rm ArT}(\vec{r})$.

The second method also assumes that $I_{\rm fore}$ corresponds to the minimum observed intensity in the target field \citep[cf.][]{Ragan+2009, Butler+2012} 
and follows a large-scale median filter approach for estimating $I_{\rm back}(\vec{r})$ in the vicinity of the absorbing structure, 
as done by \citet{Simon+2006} and \citet{Butler+2009}. 
In this method, the size of the filter needs to be sufficiently large not to be dominated by the absorbed areas, and small enough to recover the local variations of the background emission.
Here, we adopted a square median filter of size $31 \arcsec $ or $\sim$\,0.2\,pc and then derived a second estimate of the absorbing column density distribution $N_{\rm H_2}^{\rm JWST2}(\vec{r})$ using an equation similar to Equation~(\ref{equ4}). This second method assumes that the background emission has no significant fluctuations on angular scales smaller than the size of the median filter (and, as the first method, that the foreground emission is uniform). Its advantage is that it generates a high-resolution column density map directly from the mid-infrared data, without resorting to ancillary submillimeter continuum data.

The two high-resolution column density images $N_{\rm H_2}^{\rm JWST1}$, $N_{\rm H_2}^{\rm JWST2}$ 
{derived from the MIRI 7.7\,$\mu$m data} 
are shown in Figure~\ref{ngc6334_jwst_coldens_images}, where they are compared to both 
our reference ArT\'eMiS column density map $N_{\rm H_2}^{\rm ArT}$ and a ${\it Spitzer}$-based column density map 
derived from IRAC 8\,$\mu$m absorption data using method~2. 
{We similarly derived $N_{\rm H_2,\, 25.5\mu \rm{m}}^{\rm JWST1}$ and $N_{\rm H_2,\, 25.5\mu \rm{m}}^{\rm JWST2}$ column density maps 
from the MIRI 25.5\,$\mu$m data. There is an excellent linear correlation between the two pure JWST maps $N_{\rm H_2,\, 25.5\mu \rm{m}}^{\rm JWST2}$ and 
$N_{\rm H_2,\, 7.7\mu \rm{m}}^{\rm JWST2}$ (after smoothing the latter to the resolution of the former), with a Pearson correlation coefficient  $\rho \sim$\,85\%,   
even at high column densities in the vicinity of the filament crest. 
The $N_{\rm H_2,\, 25.5\mu \rm{m}}^{\rm JWST2}$ values are only $\sim$\,20\% lower than the $N_{\rm H_2,\, 7.7\mu \rm{m}}^{\rm JWST2}$ ones on average. 
There is also a very significant ($> 130\sigma$), albeit weaker ($\rho \sim$\,53\%), correlation between $N_{\rm H_2,\, 7.7\mu \rm{m}}^{\rm JWST2}$ and $N_{\rm H_2,\, 7.7\mu \rm{m}}^{\rm JWST1}$ 
in the vicinity of the filament crest, $N_{\rm H_2,\, 7.7\mu \rm{m}}^{\rm JWST2}$ being $\sim 50\% $ weaker on average than $N_{\rm H_2,\, 7.7\mu \rm{m}}^{\rm JWST1}$. 
These comparisons provide an idea of the uncertainty in our JWST-based column densities, which we estimate to be  $\sim$\,20--50\%.
}

A potential limitation of column density maps derived from absorption is saturation at the highest column densities:  
when $I_{\rm obs}(\vec{r})$ approaches $I_{\rm fore}(\vec{r})$  and is consistent with the foreground emission 
within the noise of the data, Equation~(\ref{equ2}) diverges and the results become very uncertain. 
However, the mid-infrared emission observed with JWST near 
the crest of NGC~6334M 
(see blue curve in Figure~\ref{ngc6334_jwst_coldens_images}d) is anticorrelated with 
the column density independently measured with {\it Herschel}/APEX,  
{as can be seen by directly comparing Figure~\ref{ngc6334_jwst_images}a to Figure~\ref{ngc6334_jwst_coldens_images}a.
This (non-linear) anticorrelation is statistically very significant (at confidence levels of $\sim$\,$10\sigma$ at $7.7,\mu \rm{m}$ and $\sim$\,$6\sigma$ at $25.5\,\mu \rm{m}$, 
according to Spearman correlation tests), suggesting that saturation effects are not an issue here, 
except possibly at the location of dense cores or clumps along the filament (see Appendix~\ref{app:saturation} for further discussion).}

\section{Radial structure analysis}
\label{sec:rad_prof}

\begin{figure}[h!]
\centering 
\resizebox{0.99\hsize}{!}{\includegraphics[angle=0]{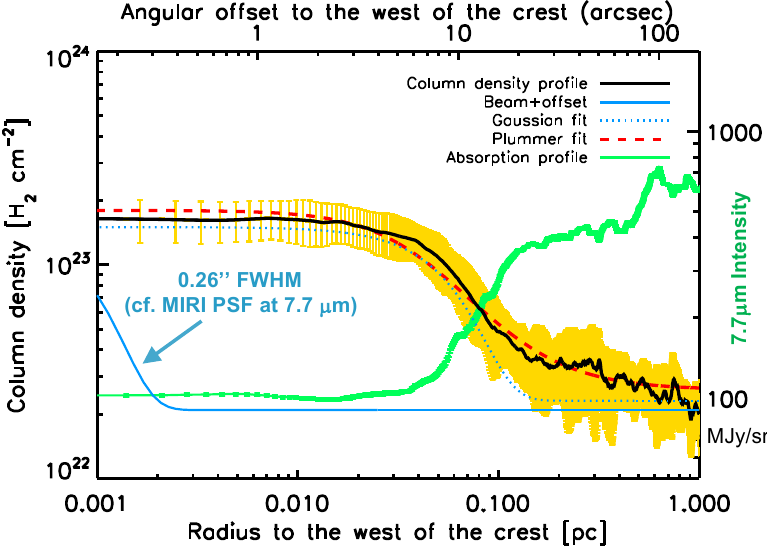}}
\caption{Median transverse column density profile of the NGC~6334M filament derived from JWST 7.7\,$\mu$m absorption data. 
Yellow error bars mark the dispersion of the distribution of radial profiles. 
Plummer and Gaussian fits (dashed red and dotted blue curves) yield a median HP diameter of $0.12 \pm 0.02$\,pc for this filament. 
}
\label{rad_profile}
\end{figure}

In Figure~\ref{rad_profile}, we present the median radial column density profile of the NGC~6334M filament 
(black curve with yellow error bars) 
as derived on the west of the filament crest from the high-resolution column density map $N_{\rm H_2}^{\rm JWST1}(\vec{r})$ shown in 
Figure~\ref{ngc6334_jwst_coldens_images}d. Also shown for comparison as a green curve is the median radial intensity profile observed across 
the filament in the JWST 7.7\,$\mu$m image of Figure~\ref{ngc6334_spitzer_jwst_images}b. 
To construct these radial profiles, we used the filament crest determined by \citet{Andre+2016} with the DisPerSE algorithm  \citep{Sousbie2011} 
on the APEX/ArTéMiS 350\,$\mu$m map of NGC~6334. We then took perpendicular cuts at each pixel of both 
the $N_{\rm H_2}^{\rm JWST1}(\vec{r})$ and the JWST 7.7\,$\mu$m map along the filament crest. 
Figure~\ref{rad_profile} displays the western part of the resulting median profiles in log-log format. 

Following \citet{Arzoumanian+2011} and \citet{Palmeirim+2013}, we then characterized  
the radial density structure of the filament by fitting both sides (west and east) 
of the observed transverse column density profiles 
with a Plummer-like model function of the form: 
\begin{equation}
N_p(r) = \frac{N_0}{\left[1 + \alpha_p\,(r/R_{\rm HP})^2\right]^{\frac{p-1}{2}}} + N_{\rm bg}, 
\label{equ5}
\end{equation}
\noindent
where $R_{\rm HP}$ is the half-power (HP) radius of the model profile,  
$p$ the power-law exponent of the corresponding density profile at radii much larger than $R_{\rm HP}$,
$\alpha_p = 2^{\frac{2}{p-1}}-1$, 
$N_0$ is the central column density, and $N_{\rm bg}$ the column density of the background cloud 
in the immediate vicinity of the filament\footnote{{Here, in practice, $N_{\rm bg}$ was initially estimated as the minimum column density 
along the (one-sided) radial profile being fitted (cf. Figure~\ref{rad_profile}) and then left as a free parameter of the Plummer fit. 
As pointed out by \citet{Schuller+2021}, we stress that the exact value of $N_{\rm bg}$ has little impact on the derived $D_{\rm HP}$ diameter
for a high-contrast filament such as NGC~6334M (for which $N_{\rm bg} \ll N_0$).}}.
For the western portion of the median radial column density profile, the best-fit model profile (dashed red curve 
in Figure~\ref{rad_profile}) has the following parameters: 
half-power diameter \hbox{$D_{\rm HP} = 2 \times R_{\rm HP} = 0.12 \pm 0.02$\,pc} (taking beam convolution into account), 
$p = 2.2 \pm 0.3$, $N_0 = (1.5\pm0.2) \times 10^{23}\, \rm{cm}^{-2}$, 
and $N_{\rm bg} = (2.5\pm0.5) \times 10^{22}\, \rm{cm}^{-2}$.  
The HP diameter $D_{\rm HP}$ derived from Plummer fitting is in good agreement 
with the half-diameter $hd \equiv 2\,hr = 0.13$\,pc estimated in a simpler way, without any fitting,  
as twice the half-radius $hr$ where the background-subtracted column density profile drops to 
half of its maximum value on the filament crest \citep[cf.][]{Arzoumanian+2019}.

Integrating the best-fit column density profile 
over radius up to $±\pm D_{\rm HP}$, we  estimate of the median mass per unit length 
for the inner portion of the filament to be $M_{\rm line} \approx 500\, M_\odot \, {\rm pc}^{-1} $. 
For a cylindrical model filament with a Plummer-like column density profile following Equation~(5), 
the underlying volume density profile has a similar form:
\begin{equation}
n_p(r) = \frac{{n}_{0}}{\left[1+\alpha_p\,\left({r/R_{\rm HP}}\right)^{2}\right]^{p/2}}, 
\label{equ6}
\end{equation}
where $n_{0}$ is the central volume density of the filament. 
The latter is related to the projected central column density $N_{0}$ by the simple relation, 
$n_{0} = \sqrt{\alpha_p}\,N_{0}/(A_p\, R_{\rm HP}) $, where 
$A_p = \frac{1}{\cos\,i} \times B\left(\frac{1}{2},\frac{p-1}{2}\right) $ 
is a constant factor which depends on the inclination angle $i$ of the filament 
to the plane of sky, and $ B$ is the Euler beta function \citep[cf.][]{Palmeirim+2013}. 
Assuming $i = 0\degr $, the mean central density of the filament segment imaged with JWST 
is estimated to be $n_0 \sim 5 \times 10^5\, {\rm cm}^{-3} $.

\begin{figure}[h!]
\centering
\resizebox{0.99\hsize}{!}{\includegraphics[angle=0]{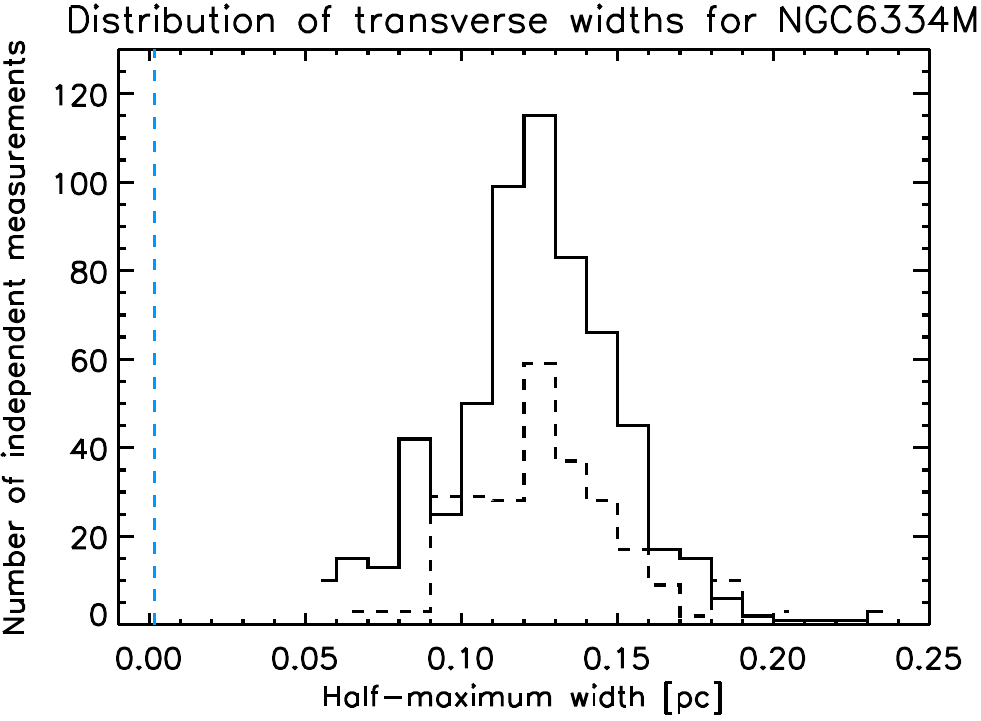}}
\caption{Distribution of individual width measurements made at beam-spaced positions 
along the NGC6334 filament using the high-resolution column density map $N_{\rm H_2}^{\rm JWST1}$ 
of Figure~\ref{ngc6334_jwst_coldens_images}d. 
The solid histogram shows the distribution of one-sided $hd$ widths obtained from 609 independent measurements 
on the west of the filament crest. The dashed histogram is the distribution of two-sided $hd$ widths for 262 independent measurements 
on either side (west $+$ east) of the crest. 
A blue dashed line marks the spatial resolution ($\sim 1.6 \times 10^{-3}\,$pc) of the JWST data at 7.7\,$\mu$m.
}
\label{histo_widths}
\end{figure}

To quantify the magnitude of the fluctuations in the HP diameter of the filament along its crest, we display 
the distribution of individual width measurements made at beam-spaced positions along the filament crest 
in Figure~\ref{histo_widths}.
In the southern portion of the filament, reliable width measurements are possible on either side (west and east)
of the filament crest. The dashed histogram in Figure~\ref{histo_widths} represents the distribution 
of two-sided half-diameters $hd_{\rm2-sided} =  (hd_{\rm west} + hd_{\rm east})/2 $ measured in this 
part of the filament. 
In the northern portion of NGC~6334M  
in Figure~\ref{ngc6334_jwst_coldens_images}, 
the presence of  two massive protostellar clumps with bright mid-infrared emission at the northeastern edge
of the field imaged with JWST (NGC$\,$6334$\,$I and I-N -- see \citealp{Sandell2000}) 
prevents us from performing a meaningful radial profile analysis on the eastern side of the filament crest, 
but reliable measurements remain feasible on the western side.
The solid histogram in Figure~\ref{histo_widths} displays the distribution 
of one-sided half-diameters $hd_{\rm west}$ measured on the northwestern side of the crest for the whole filament. 
Both distributions have a median value of 0.125\,pc and an interquartile range of 0.03\,pc (corresponding to 0.02\,pc 
in Gaussian statistics).
{Note that Figure~\ref{histo_widths} includes width measurements at all positions, 
irrespective of the presence of dense cores along the filament crest.  There is also no significant dependence of 
the measured $hd$ widths on the crest column density, with a Pearson correlation coefficient $|\rho|< 10\%$ 
(deviating from that expected for two uncorrelated Gaussian random variables by only $\sim 1\sigma$).}

While the above analysis is {\it a priori} somewhat dependent on the definition of the filament crest, 
previous work \citep[e.g.,][]{Schuller+2021} has shown that this dependence is weak. 
Indeed, different filament-finding algorithms tend to yield consistent results for high-contrast filaments 
such as NGC~6334M.
However, since a given filament may split up in several subfilaments, each of them with their own crest, 
when observed at higher resolution and/or sensitivity, 
{we also performed a characterization of the radial 
structure of NGC~6334M based on the autocorrelation function (ACF) of the JWST data.
This ACF analysis, presented in Appendix~\ref{app:ACF}, is completely independent of any detailed crest definition. 
Our ACF results are entirely consistent 
with the filament width estimates derived  from direct measurements of the radial 
profiles (see Figure~\ref{ACF_profiles} in Appendix~\ref{app:ACF} and Table~\ref{table_widths} below).}

\begin{figure}[!h]
\centering
\resizebox{0.99\hsize}{!}{\includegraphics[angle=0]{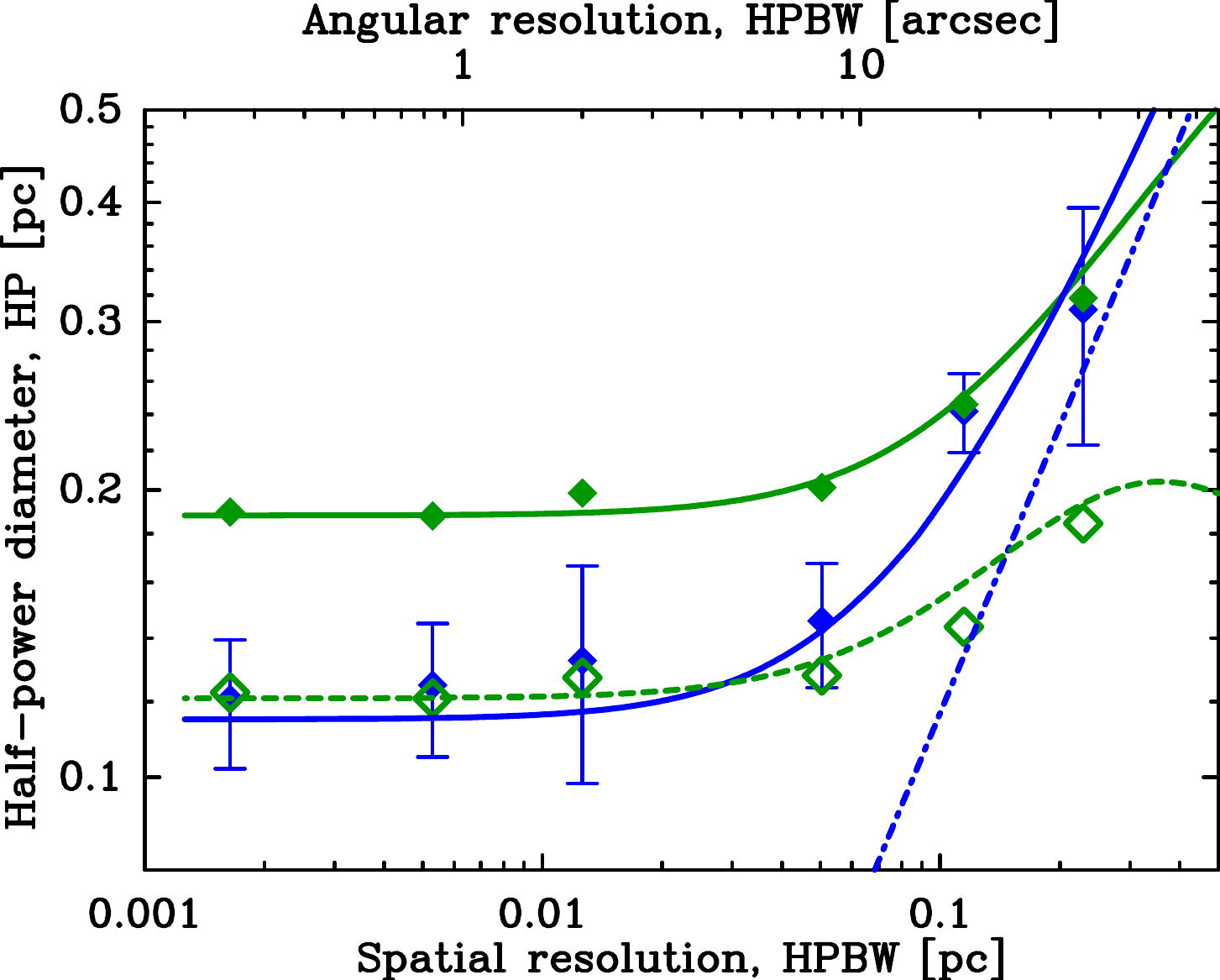}}
\caption{Convergence of HP width measurements at high resolution for the NGC6334M filament: 
Mean $hd$ width (blue filled diamonds and error bars, 
corresponding to the standard deviations of measured values along the filament), 
equivalent FWHM of ACF profile in the field mapped with JWST (green open diamonds), 
and equivalent FWHM of ACF profile in the extended field mapped with {\it Spitzer} 
and APEX/ArT\'eMiS (green filled diamonds) 
as a function of HPBW spatial resolution. 
The measurements were made in column density maps with angular resolutions 
ranging from $0.26\arcsec$ to $36.3\arcsec$: $N_{\rm H_2,\, 7.7\mu \rm{m}}^{\rm JWST1}$, 
$N_{\rm H_2,\, 25.5\mu \rm{m}}^{\rm JWST1}$, $N_{\rm H_2,\, 8\mu \rm{m}}^{Spitzer1}$, 
$N_{\rm H_2}^{\rm ArTeMiS}$,  $N_{\rm H_2,\, {\rm high}}^{Herschel}$, and $N_{\rm H_2,\, {\rm std}}^{Herschel}$ (see text).
For comparison, curves show the theoretical expectations for a cylindrical Plummer-like filament with a logarithmic density 
slope $p = 2$  and intrinsic HP width $D_{\rm HP} = 0.12$\,pc:  apparent $hd$ width (in solid blue) and 
ACF equivalent FWHM in the JWST mask (dashed green) and the {\it Spitzer} mask (solid green).
The dash-dotted blue line shows the apparent $hd$ width for a pure power-law model filament with $p = 2$.
}
\label{n6334_conv_test}
\end{figure}

\section{Convergence test for the NGC~6334 filament and comparison with other filaments}
\label{sec:conv_test}

\begin{figure*}
\centering
\resizebox{0.98\hsize}{!}{\includegraphics[angle=0]{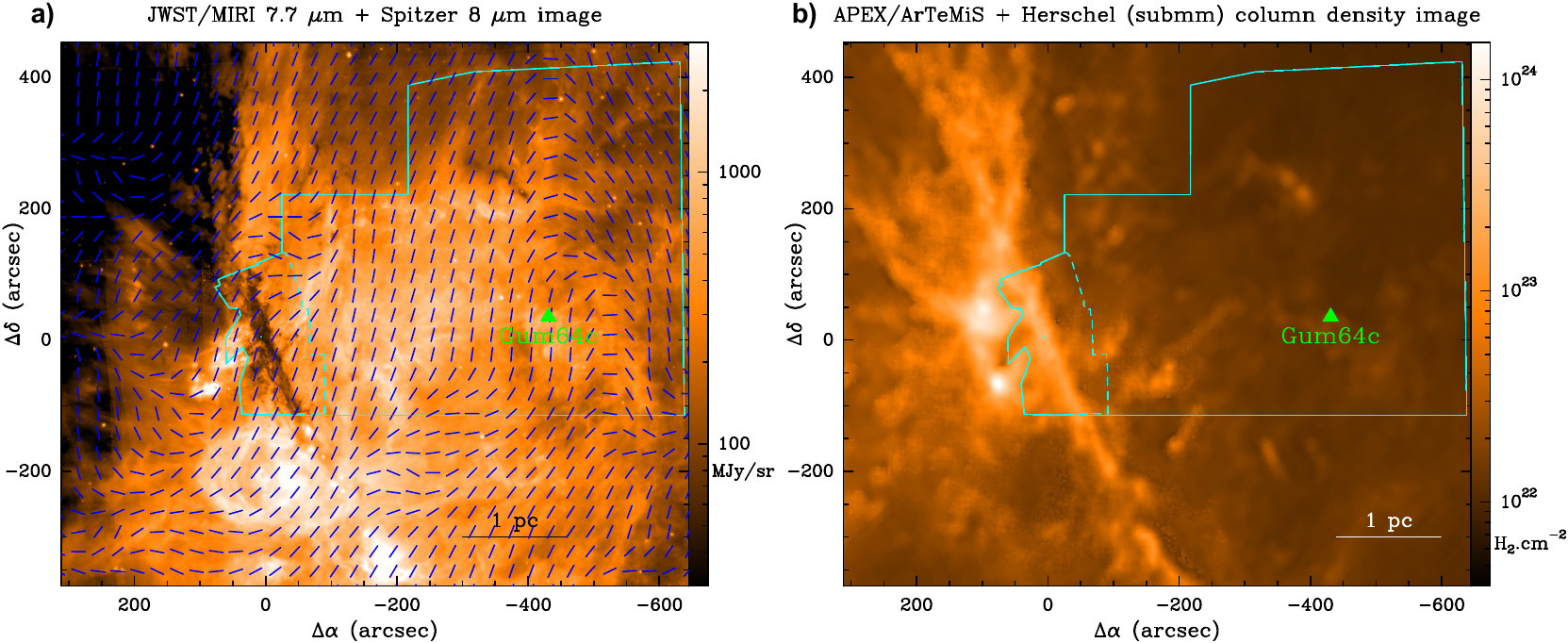}} 
\caption{Multiresolution images of NGC6334: 
{\bf (a)} Combination of the JWST 7.7\,$\mu$m image with {\it Spitzer} 8\,$\mu$m data; 
the resolution of this map ranges from $0.26\arcsec$, in the inner part of the field covered with JWST, 
to $2\arcsec$, in the outer part only covered with {\it Spitzer}. 
Blue segments mark the local direction of the plane-of-sky magnetic field in the ambient cloud as inferred 
from {\it Planck} polarization data  \citep[cf.][]{Arzoumanian+2021}.
{\bf (b)} Multiresolution column density map from the CAFFEINE 
project, based on a combination of APEX/ArT\'eMiS and {\it Herschel} data; the resolution 
ranges from $\sim$$8\arcsec$ at $N_{H_2}  > 4 \times 10^{22}\, {\rm cm}^{-2} $ to $18.2\arcsec$ resolution at lower column densities.
In both panels, the polygons in light blue outline the two masks considered in the convergence test described in the text 
and the green triangle marks the center of the HII region Gum\,64c \citep{Gum1955}.
}
\label{ngc6334_multi_resol_images}
\end{figure*}

It is instructive to compare the results obtained in Sect.~\ref{sec:rad_prof} at the $0.26\arcsec$ resolution of the JWST 7.7\,$\mu$m data
to a similar radial structure analysis using the JWST 25.5\,$\mu$m data at $0.84\arcsec$ resolution 
and the {\it Spitzer} 8\,$\mu$m data at $2\arcsec$ resolution, as well to results found with submillimeter emission data 
from ArT\'eMiS ($8\arcsec$ resolution -- cf. \citealp{Andre+2016}) and {\it Herschel} ($18.2\arcsec$ and $36.3\arcsec$ resolution
-- cf. \citealp{Tige+2017}). 
This is shown in Figure~\ref{n6334_conv_test}. 
As spatial dynamic range is a key factor in all structural analyses \citep[e.g.,][]{Ossenkopf+2001} 
we also consider multiresolution images 
resulting from 1) the combination of the JWST 7.7\,$\mu$m and {\it Spitzer} 8\,$\mu$m data 
(see Figure~\ref{ngc6334_multi_resol_images}a), and 
2) the combination of APEX/ArT\'eMiS and {\it Herschel} column density data 
(see Figure~\ref{ngc6334_multi_resol_images}b and \citealp{Mattern+2024} for details on the derivation 
of such multiresolution column density maps from submilimeter dust continuum data). 
This broad set of images, spanning a wide range of resolutions, 
allows us to perform a convergence test for the NGC~6334 filament, 
similar to that carried out by \citet[][see also \citealp{Panopoulou+2022}]{Andre+2022} for 
the Taurus B211/3 filament. 
In turn, we can investigate whether the filament width measurements of Sect.~\ref{sec:rad_prof} are robust or affected by 
the finite spatial resolution and/or extent of the data. 

\begin{table*}[!htbp]    
\caption{Widths of the NGC~6334 filament as estimated at various resolutions}
\label{table_widths}
\centering      
\begin{tabular}{c c c c c c}   
\hline\hline	
Map & Resol. & Resol. & $hd$ & $D_{\rm HP}$ & FWHM$_{\rm ACF}$  \\    	
        & ($\arcsec$) & (pc) & (pc) & (pc) & (pc) \\
 (1) & (2) & (3) & (4) & (5)  & (6) \\
\hline                        							
JWST 7.7\,$\mu$m & 0.26 & 0.0016 & $0.20 \pm 0.03$  & -- & 0.10--0.25  \\		
JWST 7.7\,$\mu$m $N_{\rm H_2}^{\rm JWST1}$ & 0.26 & 0.0016 & $0.13 \pm 0.02$  & $0.12 \pm 0.02$ & 0.125--0.18  \\
JWST 25.5\,$\mu$m & 0.84 & 0.0053 & -- & -- & 0.13--($\dag$)  \\
JWST 25.5\,$\mu$m $N_{\rm H_2}^{\rm JWST1}$ & 0.84 & 0.0053 & $0.125 \pm 0.02$  & $0.13 \pm 0.02$ & 0.12--($\dag$) \\
Spitzer 8\,$\mu$m & 2.0 & 0.0126 & -- & -- & 0.11--0.26 \\
Spitzer 8\,$\mu$m $N_{\rm H_2}^{\rm JWST1}$  & 2.0 & 0.0126 & $0.13 \pm 0.03$ & $0.12 \pm 0.02$ & 0.125--0.18\\
ArT\'eMiS$+${\it Herschel} $N_{\rm H_2}$ & 8.0 & 0.050 & $0.14 \pm 0.02$ & $0.11 \pm 0.03$ & 0.13--0.18 \\
{\it Herschel} hires $N_{\rm H_2}$ & 18.2 & 0.115 & $0.24 \pm 0.04$ & $0.16 \pm 0.05$ & 0.14--0.22 \\
{\it Herschel} $N_{\rm H_2}$ & 36.3  & 0.229 & $0.31 \pm 0.06$ & $0.29 \pm 0.09$ & 0.18--0.29 \\
\hline
\end{tabular}
\tablecomments{Col.~{\bf(4)}: Mean and standard deviation of the distribution of  half-diameters $hd \equiv 2\,hr$ estimated without any fitting along the filament.
Col.~{\bf(5)}: ``Deconvolved'' half-power diameter $D_{\rm HP}$ derived by fitting the observed mean radial profile of the filament 
with a model corresponding to Equation~(5), convolved with a Gaussian beam of HPBW given in Col.~{\bf(2)} (in arcsec) and Col.~{\bf(3)} (in pc).  
Col.~{\bf(6)}: Equivalent mean FWHM diameter derived from analysis of the transverse ACF profile as $W_{\rm HP}^{\rm ACF}/\sqrt{2}$ (see {Appendix~\ref{app:ACF}}). 
The first FWHM$_{\rm ACF}$ value corresponds to the JWST mask, the second value to the {\it Spitzer} mask ($\dag$:~No good multiresolution JWST--{\it Spitzer} image could be constructed at 25.5\,$\mu$m, preventing any ACF estimate in the  {\it Spitzer} mask).
}
\end{table*}

Figure~\ref{n6334_conv_test} and Table~\ref{table_widths} summarize 
the width estimates we made at various resolutions, 
through both direct measurements of the filament's radial profiles (cf. Figure~\ref{rad_profile}) and analysis 
of the corresponding ACF images and transverse cuts ({see Figures~\ref{autocorrel_images} and \ref{ACF_profiles} in Appendix~\ref{app:ACF}}). 
Since, as {explained in Appendix~\ref{app:ACF}}, 
the ACF characterizes an entire input image, 
it is important to restrict the calculation of the ACF function ({Equation~\ref{equ6}}) to relevant portions of the input image. 
The two masks we considered in our multiresolution analysis 
are outlined as light-blue polygons in the wide-field images 
shown in Figure~\ref{ngc6334_multi_resol_images}. 
The smaller mask corresponds to the portion of the field imaged with JWST after excluding the two massive protostellar clumps 
NGC$\,$6334$\,$I and I-N to the east of the filament (dashed polygon in Figure~\ref{ngc6334_multi_resol_images}).
The larger mask includes the first mask but extends significantly father to the west of the filament, 
out to the edge of the local tile image from the {\it Spitzer} GLIMPSE 8\,$\mu$m survey 
(see solid polygon in Figure~\ref{ngc6334_multi_resol_images}). 
We will refer to these two masks as the JWST and {\it Spitzer} masks, respectively. 
As can be seen in Figure~\ref{n6334_conv_test}, the $hd$ widths derived from the filament's radial profiles (blue diamonds) and 
both sets of ACF width measurements (open and filled green symbols based on the JWST and {\it Spitzer} masks, respectively) 
are essentially independent of 
spatial resolution for HPBW\,$\la 0.05\,$pc and increase markedly with resolution for HPBW\,$>0.1\,$pc. 
In particular, the results obtained from the 7.7 and 25.5\,$\mu$m JWST data are remarkably similar, 
despite a factor of $> 3$ difference in angular resolution. 
Moreover, the measurements are consistent with the dependence with resolution expected for a cylindrical Plummer-like model filament
following Equation~(5) with $p = 2$  and $D_{\rm HP} = 0.12$\,pc (blue and green curves in Figure~\ref{n6334_conv_test}).

\begin{figure}[!htbp]
\centering
\resizebox{0.99\hsize}{!}{\includegraphics[angle=0]{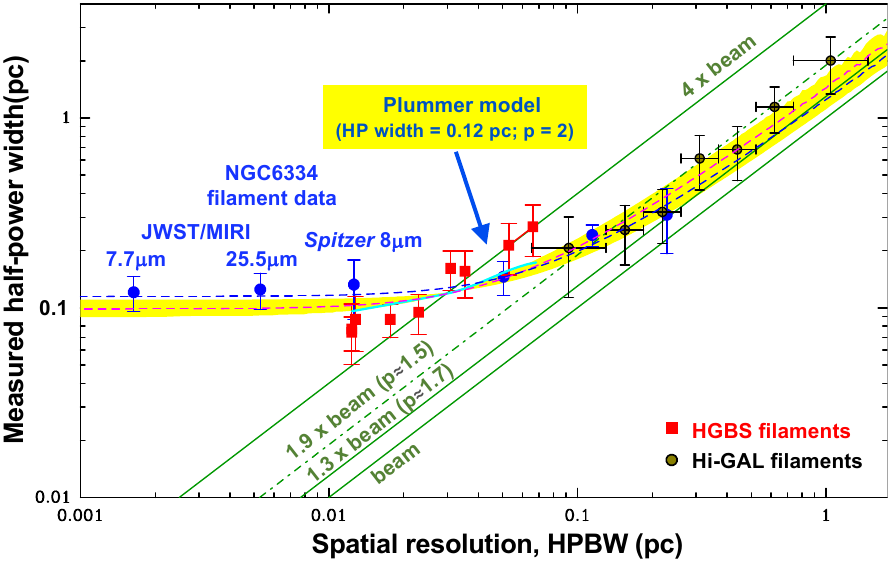}} 
\caption{Mean apparent half-power (HP) width vs. spatial resolution for the NGC6334M filament 
based on JWST, {\it Spitzer}, ArTéMiS, and {\it Herschel} (blue symbols -- see text) 
and for {\it Herschel} filaments from both the 
HGBS \citep[red squares;][]{Arzoumanian+2019} and Hi-GAL \citep[black circles;][]{Schisano+2014} surveys. 
The vertical error bars correspond to $\pm$ the standard deviations of 
measured widths in each HGBS cloud or Hi-GAL resolution bin. 
The solid light blue curve shows the results of the convergence test performed 
by \citet{Andre+2022} for the Taurus filament.  
The dashed blue curve 
shows the theoretical expectation for a cylindrical Plummer-like filament
with a logarithmic density slope $p = 2$  and intrinsic HP width $D_{\rm HP} = 0.12$\,pc 
(consistent with the radial profile analysis in Sect.~\ref{sec:rad_prof} and Fig.~\ref{rad_profile}); 
the dashed pink curve is the same for a Plummer model with $p = 1.7$ and  $D_{\rm HP} = 0.1$\,pc.
The variations in expected HP width induced by variations of the 
$p$ index in Equation~(\ref{equ5}) between 1.5 and 2.5 are displayed by a yellow shading.
}
\label{width_vs_resol}
\end{figure}

The results of our convergence test are further illustrated in Figure~\ref{width_vs_resol}, 
which compares our present width measurements for NGC~6334M (blue filled circles) 
with the findings of \citet{Andre+2022} for the Taurus B211/3 filament (light blue curve), 
the mean filament widths found by \citet{Arzoumanian+2019} in 8 nearby HGBS clouds (red filled squares) 
and by \citet{Schisano+2014} for a sample of Hi-GAL filaments in the Galactic Plane 
(see \citealp{Andre+2022} for details on the latter data points). 
It can be seen that {apparent filament widths, i.e., measured widths with no attempt of any beam deconvolution,} 
tend to converge toward 
a common value of $\sim$\,0.1\,pc when this scale is resolved by a factor of $\sim$\,10. 
Furthermore, a single Plummer-like model with $D_{\rm HP}^{\rm int} \approx 0.1$\,pc and $p \approx 1.7$--2  
(see dashed pink and blue curves in Figure~\ref{width_vs_resol}) 
can account for all of the measurements reasonably well. 
This suggests that dusty molecular filaments do have a typical HP width $\sim$0.1\,pc 
and typical power-law wings with $p \sim 1.5$--2, but that the flat inner portion of their 
column density profiles is unresolved by {\it Herschel} at the distances $d \sim 1$--3\, kpc 
of the Hi-GAL filaments in Figure~\ref{width_vs_resol}. 

\section{Discussion and conclusions}
\label{sec:concl}

Thanks to our JWST/MIRI data, we could construct column density maps of the NGC~6334M filament with 
unprecedented resolution and dynamic range (cf. Sect.~\ref{sec:coldens} and Figure~\ref{ngc6334_jwst_coldens_images}). 
These maps have a spatial dynamic range comparable to the {\it Herschel}-based column density maps
of nearby molecular clouds produced as part of the HGBS survey \citep[e.g.,][]{Palmeirim+2013,Konyves+2015}, but probe an order-of-magnitude 
smaller scales. While {\it Herschel} maps of 
clouds such as Taurus \citep[][]{Palmeirim+2013} sample angular scales between $\sim$\,$18\arcsec$
and $\sim$\,2$\degr$--3$\degr$, corresponding to spatial scales from $\sim$\,0.012\,pc and $\sim$\,5--7\,pc at $d \sim 140$~pc, 
our present column density maps $N_{\rm H_2}^{\rm JWST1}$ and $N_{\rm H_2}^{\rm JWST2}$ of NGC~6334M, shown in panels d and c of 
Figure~\ref{ngc6334_jwst_coldens_images} respectively, sample angular scales between $\sim$\,$0.26\arcsec$ 
and $\sim$\,100--200$\arcsec$, or spatial scales from $\sim$\,$1.6\times10^{-3}\,$pc to $\sim$\,0.65--1.3\,pc.
This allows us to set new constraints on the structure of the NGC~6334 
molecular cloud and the physics of star-forming filaments. 

\begin{figure}[!htp]
\centering
\resizebox{1.0\hsize}{!}{\includegraphics[angle=0]{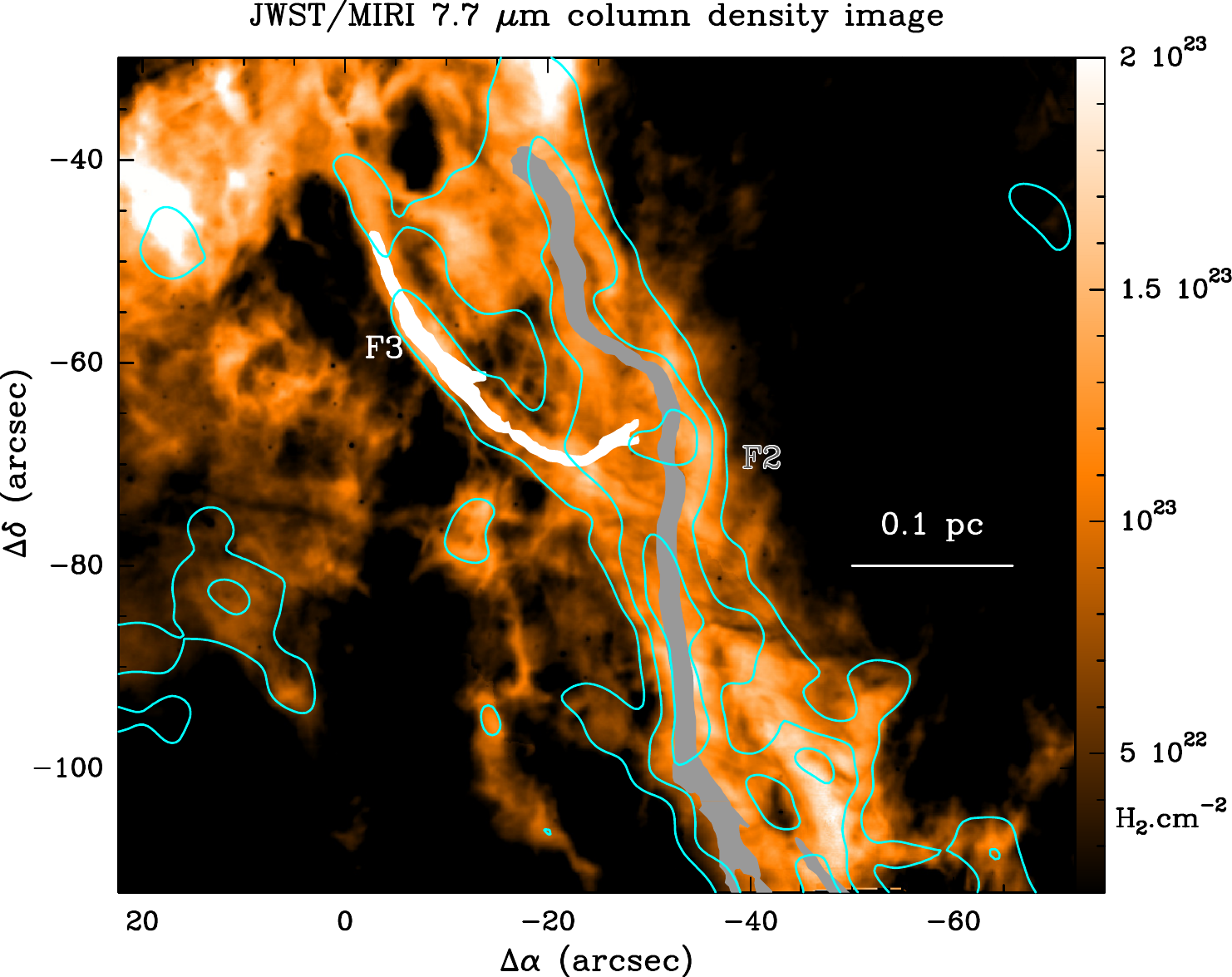}} 
\caption{Blow-up $N_{\rm H_2}^{\rm JWST2}$ column density image of the southern part of the field imaged by JWST  
with overlaid contours of N$_2$H$^+$(1--0) integrated intensity from ALMA 3\,mm observations \citep[cf.][]{Shimajiri+2019b}. 
Note how the fiber-like substructures labeled ``F2'', 
``F3" by \citet{Shimajiri+2019b} 
and displayed as thick white and grey skeleton curves here nicely show up in the JWST image.
}
\label{alma_fibers}
\end{figure}

\begin{figure}[!htbp]
\centering
\resizebox{1.01\hsize}{!}{\includegraphics[angle=0]{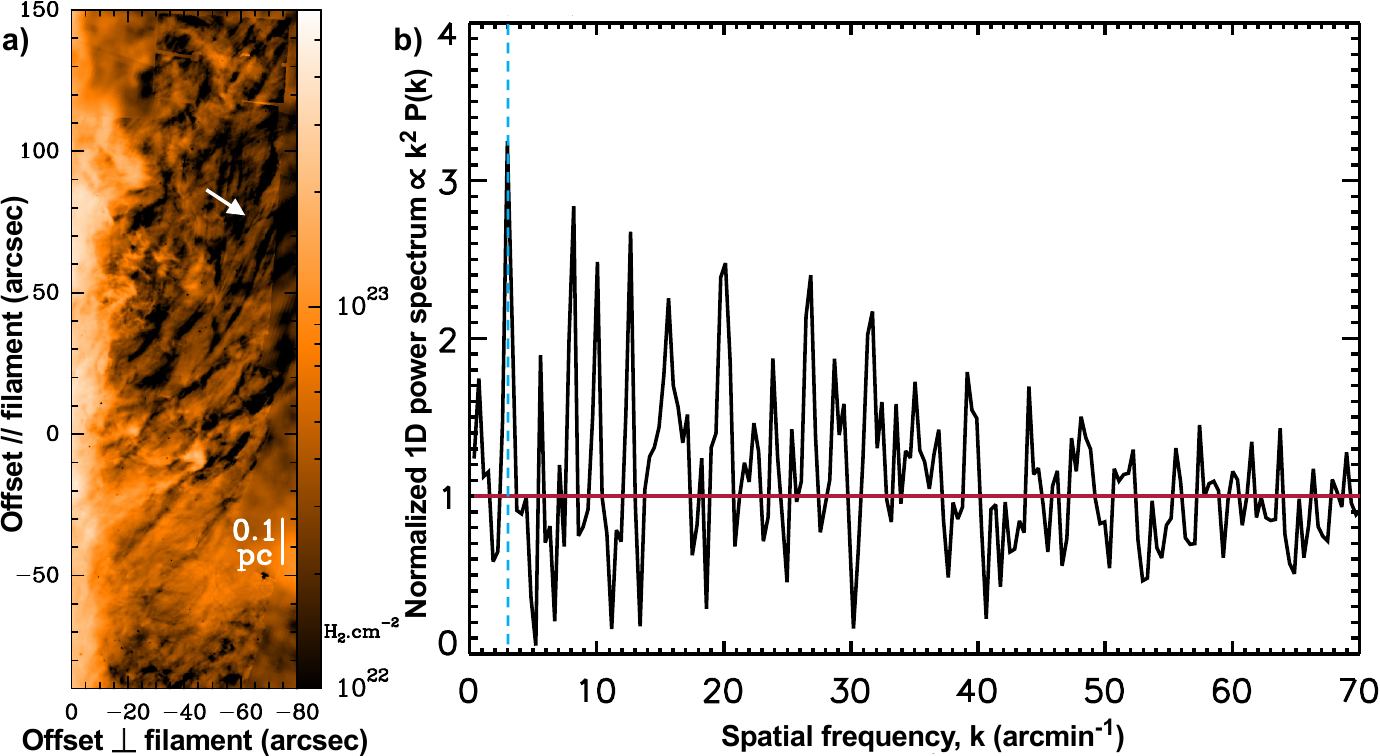}}
\caption{{\bf (a)} Close-up $N_{\rm H_2}^{\rm JWST1}$ column density map of the north-western part of the field imaged by JWST, 
showing a quasi-periodic series of side filaments connected to the main (NGC~6334M) filament. 
This close-up map has been rotated by $+21\degr$ clockwise to make the main filament parallel to the y-axis. 
 A small arrow points to one of the side filaments.
{\bf (b)}~Normalized power spectrum [$k^2\,P(k)$] of the mean column density profile 
perpendicular to the side filaments, 
scaled so as to emphasize departure from a pure power law $P(k) \propto k^{-2}$.
Note the presence of a significant ($5.5\sigma $) peak in the power spectrum at a spatial frequency of $\sim 3\pm0.3\,$arcmin$^{-1}$,  
marked by the vertical dashed line. The horizontal red line marks the best power-law fit to the observed power spectrum, which is close 
to $P_{\rm PL}(k) \propto k^{-2}$.
}
\label{striations}
\end{figure}

The JWST data reveal the presence of small substructures on scales inaccessible to {\it Herschel} or even APEX/ArT\'eMiS 
due to insufficient resolution. Two examples of small-scale features detected with JWST 
are shown in Figures~\ref{alma_fibers} and ~\ref{striations}. 
Figure~\ref{alma_fibers} illustrates the good correspondence between some of the column density fluctuations 
visible in the $N_{\rm H_2}$~JWST map(s) and the velocity-coherent substructures of dense gas identified 
with ALMA at $3\arcsec$ resolution in 
NGC~6334M using the N$_2$H$^+$(1--0) line \citep{Shimajiri+2019b}.
Even without any velocity information, at least two (``F2" and ``F3'') of the five velocity-coherent ``fiber-like'' substructures 
found with ALMA are also clearly seen by JWST. 
\citet{Shimajiri+2019b} interpret the presence of such `fiber-like'' substructures as the manifestation of accretion-driven turbulence 
into the main filament, in accordance with some numerical simulations \citep[e.g.,][]{Clarke+2017}.

Figure~\ref{striations}a emphasizes a second type of small-scale column density features visible in the JWST data, 
namely several side filaments connected to the main filament at an angle of $\sim$\,45$\degr$. 
Remarkably, the spatial distribution of these side filaments is quasi-periodic with 
a typical projected spacing\footnote{{For a random orientation of the line of sight, 
even  a high inclination angle $i \sim 60\degr $ of the side filaments to the plane of the sky 
(occurring in only $1 -{\rm sin}\,i \sim 13\%$ of the cases) would make the deprojected spacing of the side filaments 
only a factor of $\sim 2$ larger than the observed spacing.}}
in the plane of the sky of $\sim$\,$20\arcsec\pm0.2\arcsec$ or $\sim$\,$0.125\pm0.015$\,pc, 
as evidenced by the normalized power spectrum or periodogram shown in Figure~\ref{striations}b. 
These side filaments are roughly parallel to the large-scale magnetic field traced by {\it Planck} polarization data 
(see blue segments in Figure~\ref{ngc6334_multi_resol_images}a; see also \citealp{Arzoumanian+2021} ands \citealp{LiHb+2015}) 
and are reminiscent of the striations seen by {\it Herschel} in the immediate vicinity of the Taurus and Musca filaments 
\citep[cf.][]{Palmeirim+2013, Cox+2016}. They possibly represent accretion flows of ambient cloud gas (and dust) 
into the main filament, although additional kinematic information would be required to support this hypothesis. 

In addition to the above-mentioned ``fiber-like'' and side-filament structures, density fluctuations on the whole range of sampled spatial scales, 
down to the $0.26\arcsec$ or \hbox{$\sim$\,$1.6\times10^{-3}\,$pc} resolution of MIRI at 7.7\,$\mu$m, are present 
in the JWST column density data, as shown by the power spectrum of the $N_{\rm H_2}^{\rm JWST1}$ image 
(solid red curve and error bars in Figure~\ref{power_spectra}; see Appendix~\ref{app:power_spectra} for details on the derivation of the power spectrum). 
The latter is well fit by a Kolmogorov-like\footnote{The general relation between the Kolmogorov energy spectrum
 $E_{\rm K} (k) \propto k^{-5/3}$ and the velocity power spectrum in $n$-D
 space is $E_{\rm K} (k) \propto k^{n-1}\, P_{\rm v, {\it n}D}^{\rm K} (k) $ \citep[cf.][]{Elmegreen+2004}. 
In subsonic (incompressible) Kolmogorov turbulence, 
the continuity equation implies a direct proportionality between (small) velocity and density fluctuations, 
and the power spectrum of 2D column density images is 
thus expected to be $P_{\rm 2D}^{\rm K} (k) \propto k^{-\gamma_{\rm K, 2D}}$ with $\gamma_{\rm K, 2D} = 8/3 \approx 2.67$.} 
power law $P_{\rm 2D}^{\rm JWST} (k) \propto k^{-2.8\pm0.1}$ 
for spatial frequencies $k$ between $\sim$\,1\,arcmin$^{-1}$ and $\sim$\,70\,arcmin$^{-1}$ 
(dashed purple line in Figure~\ref{power_spectra}). 
Interestingly, however, the power spectrum of fluctuations derived from JWST on small $\la 0.1$\,pc scales is significantly steeper 
than the power spectrum found from {\it Herschel} column density data on larger $\sim$\,0.2--30\,pc scales 
(blue curve and error bars in Figure~\ref{power_spectra}), 
for which the best-fit power law is $P_{\rm 2D}^{Herschel} (k) \propto k^{-1.6\pm0.1}$ (dashed green line in Figure~\ref{power_spectra}). 
In the limited range of spatial frequencies well sampled by both JWST and {\it Herschel} data, the two power spectra are consistent  
with each other within uncertainties, but the marked difference in slope implies the presence of a break in the actual power spectrum 
of NGC~6334 density fluctuations at a spatial frequency $k_b \sim 2 \pm 1\,$arcmin$^{-1}$.

\begin{figure}[!htp]
\centering
\resizebox{1.0\hsize}{!}{\includegraphics[angle=0]{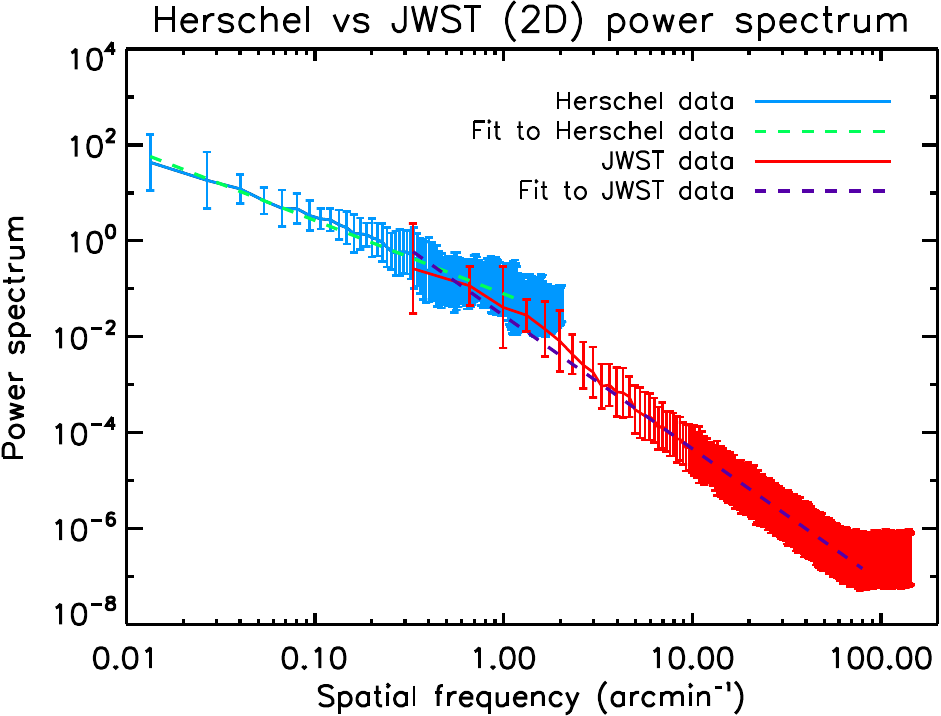}}
\caption{Comparison of the power spectrum of the JWST ($N_{\rm H_2}^{\rm JWST1}$) column density image 
of NGC~6334M on small scales (in red) with the power spectrum of 
the {\it Herschel} column density map of the broader NGC~6334 region \citep[][]{Russeil+2013,Tige+2017}, including the field imaged here with JWST, 
on larger scales (in blue). 
Note how the slopes of the two power spectra differ, indicating the presence of a significant break 
at a spatial frequency of $\sim$\,2 arcmin$^{-1}$.
}
\label{power_spectra}
\end{figure}

Despite the power-law nature of the JWST spectrum of density fluctuations, the radial structure analysis and convergence test presented  in 
Sect.~\ref{sec:rad_prof} and Sect.~\ref{sec:conv_test} demonstrate that the dominant density structure in the 
field, namely the NGC~6334M filament, has a well-defined 
HP transverse diameter of $\sim$\,$0.12\pm0.02$\,pc. 
Remarkably, the latter size scale roughly corresponds to the typical width of, and mean spacing between, 
the side filaments highlighted in Figure~\ref{striations}a. 
Even the fiber-like substructures of the main filament (Figure~\ref{alma_fibers}) have widths that are only slightly 
smaller (by a factor of $\sim$\,3--4).
Our JWST results therefore support the notion that high-contrast filamentary structures have 
a typical half-power width of $\sim$0.1\,pc, at least in the case of nearby Galactic molecular clouds, 
as argued earlier by \citet{Arzoumanian+2011,Arzoumanian+2019}.

The existence of a typical transverse size scale of $\sim$\,0.1\,pc for the dominant density structures in the NGC~6334 region
may be simply understood in terms of the broken power-law shape seen in Figure~\ref{power_spectra}
for the global power spectrum of column density fluctuations. 
The relatively shallow slope $\gamma_h = 1.6\pm0.1 < 2$ of the power spectrum at low spatial frequencies $k < k_b$, 
coupled with the steeper slope  $\gamma_j = 2.8\pm0.1 > 2$ at high spatial frequencies $k > k_b$ implies that most 
of the ``power'' (or variance) in the whole spectrum of density fluctuations resides in a relatively narrow range 
of spatial frequencies around the break frequency $k_b$. 
Indeed, the total variance $\sigma^2_N$ of column density fluctuations may be approximated as:
\begin{eqnarray}
\sigma^2_N  & = & \int_{k_{\rm min}}^{k_{\rm max}} P(k)\,2\pi k\,dk  \nonumber \\
                    & \approx & \ 2\pi P(k_b)k_b^2\, \left(\frac{1}{2-\gamma_h} + \frac{1}{\gamma_j-2}\right), 
\label{equ7}
\end{eqnarray}
where we have assumed $k_{\rm min} << k_b$ and $k_{\rm max} >> k_b$.
The break frequency corresponds to a size scale of $\sim$\,0.13--0.38\,pc, which is comparable to the typical width 
of NGC~6334M and its side filaments. 
While the structure of molecular clouds such as the NGC~6334 region is known to be largely hierarchical and self-similar \citep[cf.][]{LiHb+2015,Hacar+2023}, 
it is therefore not strictly scale free but features a characteristic scale of order $\lambda_c \sim$\,0.1--0.3\,pc, 
corresponding to the typical width of prominent star-forming filaments such as NGC~6334M.

A possible interpretation of this characteristic lengthscale and common filament width 
is that it corresponds to the sonic scale below which interstellar turbulent flows become subsonic 
and incompressible (primarily solenoidal) 
\citep{Federrath2016, Federrath+2010, Federrath+2021}. 
{As pointed out by \citet{JaupartChabrier2021}, the sonic scale is also essentially the correlation length of turbulent density fluctuations.}
Using hydrodynamic simulations of driven isothermal supersonic turbulence without self-gravity and magnetic fields, \citet{Federrath+2010} found a break 
in the power spectrum of density fluctuations roughly located at the sonic scale for compressive modes of turbulence, 
with significantly weaker density fluctuations on scales smaller than the sonic scale. 
This is reminiscent of the break seen in Figure~\ref{power_spectra}. 

The shallow power spectrum slope $\gamma_h$\,$\sim$\,$1.6$~$ < \gamma_{\rm K, 2D}$ measured with {\it Herschel} on scales $\lambda > \lambda_c$ 
is not consistent, however, with that expected from supersonic turbulence in a non self-gravitating medium. The latter is somewhat steeper than 
the Kolmogorov slope ($\gamma_{\rm turb, 2D} \ga \gamma_{\rm K, 2D}$, according to the simulations of \citealp{Federrath+2010}). 
This discrepancy likely results from strong effects of self-gravity \citep[cf.][]{Ossenkopf+2001} 
on the structure of the massive NGC~6334 cloud. 
Cloud contraction on multiple scales due to self-gravity will tend to generate a hierarchical network of centrally-condensed filaments and hubs  
with typical $\rho \propto r^{-2}$ radial density profiles \citep[e.g.,][]{Shu1977, Vazquez+2019}, and the overall cloud (or clump) itself may eventually 
acquire a $\rho \propto r^{-2}$ density distribution on average. 
At any viewing angle, the typical column density or gas surface density 
distribution around each hub will approach  $\Sigma \propto \bar{r}^{-1}$, where $\bar{r}$ denotes projected radius. 
The 2D power spectrum of such a column density distribution will be $P_{\rm 2D}^{\rm Hub} (k) \propto k^{-2}$. 
Assuming a random distribution of independent contracting hubs in the cloud, the global 2D power spectrum of the cloud or collection of hubs 
will also be close to $P_{\rm 2D}^{\rm Cloud} (k) \sim \sum_{\rm Hubs} P_{\rm 2D}^{\rm Hub} (k)  \propto k^{-2}$. 
Such a spectrum matches the NGC~6334 power spectrum observed on large scales better than a turbulent, Kolmogorov-like spectrum does. 

Most importantly, 
our JWST observations (e.g., Figure~\ref{power_spectra}) 
suggest that the power spectrum of density fluctuations induced by the combined effects 
of self-gravity and supersonic turbulence on large scales breaks down below the sonic 
scale, such as the interior of star-forming filaments 
where turbulent density fluctuations become weaker and Kolmogorov-like. 

The above interpretation of the characteristic transverse scale of dense filaments in terms of the sonic scale is, however, incomplete 
because it essentially relies on isotropic hydrodynamic turbulence and does not account for the highly anisotropic growth of self-gravitating filaments 
such as Taurus B211/3 and NGC~6334M through gas contraction and accretion along ordered magnetic field lines \citep[cf.][]{Palmeirim+2013,LiHb+2015}. 
It also does not explain how self-gravitating, thermally supercritical filaments can maintain a roughly constant inner width while evolving. 

An alternative, perhaps more promising explanation is that the filament width is linked to the ambipolar diffusion scale, 
the wavelength below which MHD waves 
cannot propagate in the mostly neutral gas of molecular clouds. For typical molecular cloud parameters, the ambipolar diffusion scale 
is on order of $\sim$\,0.1\,pc \citep[cf.][]{Hennebelle2013, Abe+2025}. On this basis, it has been proposed that the characteristic width of star{-forming} filaments
may be set by the balance between accretion-driven MHD turbulence and the dissipation of MHD waves due to ambipolar diffusion \citep[][]{HennebelleAndre2013}.

While early non-ideal MHD simulations with limited dynamic range and without self-gravity
\citep{Ntormousi+2016} 
could not establish a clear effect of ambipolar diffusion on filament widths,
more recent, higher-resolution simulations by \citet{Abe+2024, Abe+2025} have been more successful. 
The ``STORM'' (Slow-shock-mediated Turbulent flOw Reinforced by Magnetic diffusion) model introduced by \citet{Abe+2025} 
explicitly takes into account the observational finding that many, if not most, star-forming filaments appear to form and grow anisotropically 
within shock-compressed sheet-like cloud structures, through converging accretion flows along magnetic field lines 
\citep[e.g.,][and Pineda et al. 2023 for a review]{Palmeirim+2013,Arzoumanian+2018,Shimajiri+2019,Chen+2020}. These flows are mildly supersonic but sub-Alv\'enic 
and lead to slow shocks which act to confine the filaments to finite radii. Due to the slow-shock MHD instability the shock fronts are corrugationally 
unstable, and ambipolar diffusion allows the gas to flow across the magnetic field around the shock, generating dense blobs that transfer momentum 
and drive internal turbulence. In essence, the STORM mechanism thus converts the kinetic energy of accretion flows into anisotropic turbulence 
in the filaments, providing an effective pressure gradient that helps to maintain a roughly constant filament width. 
The characteristic length scale over which the dense blobs develop at the surface of accreting filaments corresponds to the most unstable scale 
of the slow-shock instability, which is proportional to the ambipolar diffusion scale \citep{Abe+2024}. 

\begin{figure}[!htbp]
\centering
\resizebox{1.01\hsize}{!}{\includegraphics[angle=0]{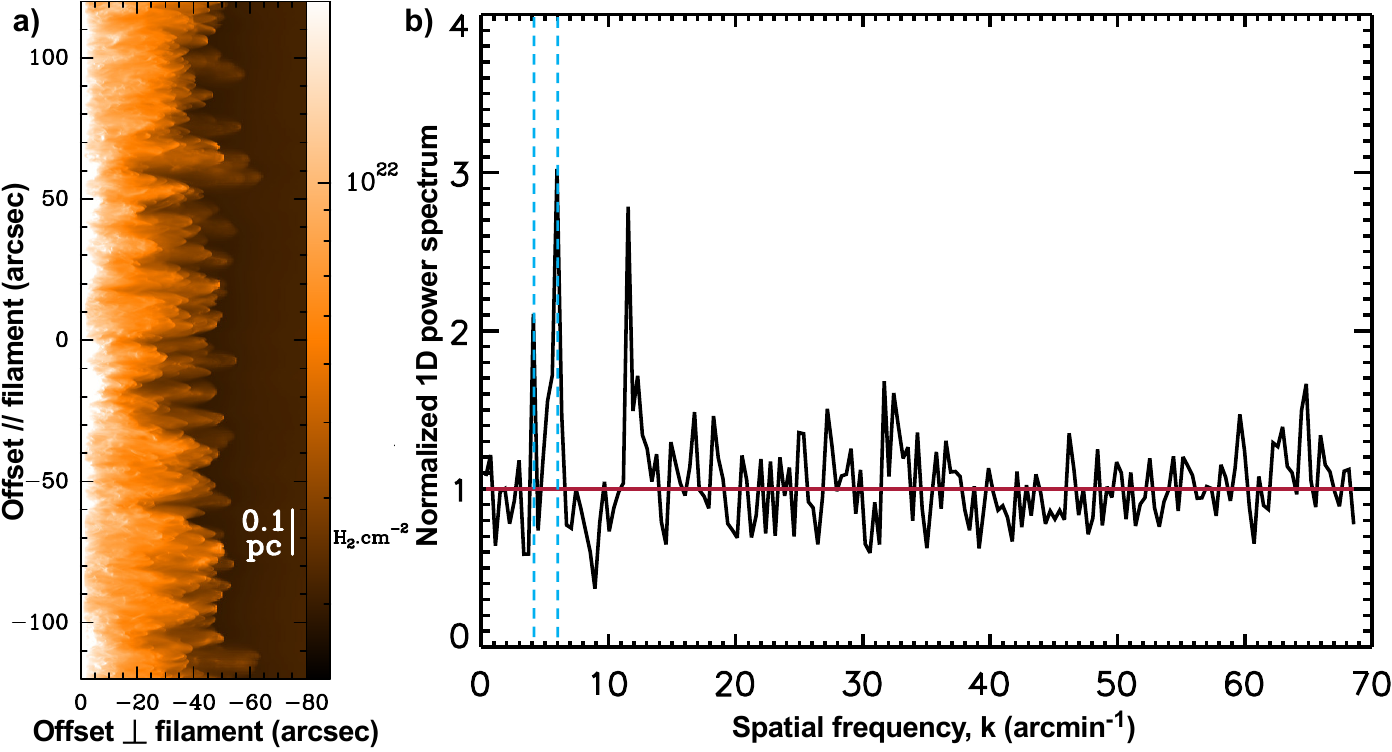}}
\caption{Similar to Figure~\ref{striations} for the STORM numerical simulations of \citet{Abe+2025}, 
assuming a simulated filament at the same distance $d = 1.3$\,kpc as NGC~6334 for direct comparison.
{\bf (a)} Close-up column density map of the simulated filament, focusing on the quasi-periodic pattern of 
finger features perpendicular to the filament.
{\bf (b)}~Mean 1D power spectrum of the map in panel {\bf a)} along the main axis of the filament, 
normalized by the overall Kolmogorov-like shape of the power spectrum. 
Note the presence of three significant peaks in this normalized power spectrum at spatial frequencies 
of $\sim 4\pm0.3\,$arcmin$^{-1}$ ($4.8\sigma $) , $\sim 6\pm0.3\,$arcmin$^{-1}$ ($8.8\sigma $), 
and $\sim 12\pm0.3\,$arcmin$^{-1}$ ($7.8\sigma $). The third peak is located at the third and second harmonics 
of the first two spatial frequencies. 
The latter two fundamental frequencies, marked by vertical dashed lines, correspond to characteristic spacings of $\sim 0.09\,$pc 
and $\sim 0.06\,$pc, both comparable to the HP width of the simulated filament \citep[cf.][]{Abe+2025}.
}
\label{simus}
\end{figure}

The simulations presented by \cite{Abe+2025} show that the STORM mechanism allows massive self-gravitating filaments to retain 
a half-power width of $\sim$\,0.1\,pc as observed here for NGC~6334M for at least $\sim$\,1\,Myr. 
The latter is a factor of $> 20$ longer than the free-fall timescale in the flat central portion of NGC~6334M.
Two additional merits of the STORM model are 1) that it generates Kolmogorov-like turbulence in the interior of filaments, in agreement with 
the power spectrum of density fluctuations found here with JWST  (cf. red curve in Figure~\ref{power_spectra}) 
and 2) that it leads to a quasi-periodic corrugation pattern of blob fingers or hail features, parallel to the magnetic field lines, 
at the surface of dense filaments (Figure~\ref{simus}), 
which resembles the quasi-periodic distribution of side filaments observed in the immediate vicinity of NGC~6334M 
with JWST (Figure~\ref{striations}). 
The main difference between the simulated and the observed pattern is the pronounced angle of $\sim$\,45$\degr$ between the present side filaments 
and the normal direction to the main filament in the observations. 
{First, the side filaments may be  intrinsically perpendicular to the main filament as in the simulations and the observed $\sim$\,45$\degr$ angle 
may result from a projection effect if the plane defined by the side filaments and the main filament is highly inclined to the plane of the sky, by say $\sim 70\degr$.
Indeed, two vectors that are perpendicular in 3D can be observed to make a wide variety of angles 
when seen in 2D projection \citep[cf. Appendix C.3 of][]{PlanckXXXV2016}. Assuming a random orientation of the line of sight with respect to the filaments, 
the observed configuration is not the most likely situation, however, and would occur only in $< 30\%$ of the cases in that hypohesis.}
{Another, more likely explanation is that the observed $\sim$\,45$\degr$ angle results 
from the effect} of several expanding HII bubbles in the immediate vicinity of NGC~6334M 
\citep[cf.][see also the central position of Gum~64c in Figure~\ref{ngc6334_multi_resol_images}]{Tahani+2023,Neupane+2024}. 
There is evidence from extensive CO(3--2) mapping with APEX that these bubbles have shaped the large-scale distribution 
and velocity structure of the molecular gas in the NGC~6334 complex \citep{Neupane+2024} and 
presumably the geometry of the ambient magnetic field (blue segments in Figure~\ref{ngc6334_multi_resol_images}a) as well. 

Regardless of the detailed physical mechanism explaining our results, the observational existence of 
a characteristic transverse correlation length and typical HP width $\sim$\,0.1\,pc for star-forming filaments in nearby ($d \la 1.5\,$\,kpc) molecular clouds 
has important implications for our understanding of the global statistical outcomes of the star formation process, namely the star formation rate (SFR) 
or efficiency (SFE) and the stellar IMF on cloud -- and perhaps even galactic -- scales. 

First, given the typical filament width $\sim$\,0.1\,pc, the thermally critical mass per unit length $M_{\rm line, crit} \sim 16\, M_\odot \, {\rm pc}^{-1} $
of nearly isothermal molecular filaments at $T \sim 10\,$\,K (see Sect.~\ref{sec:intro}) sets a natural volume density or column density transition at 
$n_{H_2} \sim 2 \times 10^4\, {\rm cm}^{-3} $ or \hbox{$N_{H_2} \sim 7.5 \times 10^{21}\, {\rm cm}^{-2} $} below which core/star formation is expected 
to be very inefficient and above which the star formation efficiency (${\rm SFE} \equiv {\rm SFR}/M_{\rm gas}$) may be expected to be 
approximately constant at a typical value ${\rm SFE_{dense}} \sim 5 \times 10^{-8}\, {\rm yr}^{-1}$ \citep[cf.][]{Andre+2014}. 
This is in agreement with observations suggesting a quasi-universal star formation law in the dense gas of giant molecular clouds (GMCs) within galaxies 
\citep{Gao+2004, Lada+2012, Evans+2014, Shimajiri+2017, Mattern+2024}. 

Second, the typical filament width introduces a characteristic fragmentation length scale in molecular clouds. 
For example, linear fragmentation models of cylindrical filaments
predict that the typical core spacing should directly scale with the filament diameter \citep[e.g.][]{Inutsuka+1992}.
Coupled with the thermally critical mass per unit length of molecular filaments, this typical fragmentation length scale
leads to a characteristic prestellar core mass in transcritical filaments, which may account for the broad peak of the prestellar core mass function 
(CMF -- cf. \citealp{Konyves+2015}) and, by extension, the ``base'' of the stellar IMF. 
More generally, the characteristic filament width results in a strong correlation between the line mass of a self-gravitating filament and 
the typical mass of the prestellar cores that may form by gravitational fragmentation of this filament \citep[cf.][]{Shimajiri+2019b, Shimajiri+2023}. 
This implies that the high-mass, power-law tail of the CMF and IMF may be partly inherited from the observed Salpeter-like distribution 
of supercritical filament masses per unit length \citep{Andre+2019}.

Our JWST study of the NGC~6334M region illustrates the unique power of the MIRI instrument in providing mid-infrared absorption images of infrared-dark clouds 
and filaments with unprecedented quality. These images  set new 
constraints on the small-scale structure of the molecular interstellar medium and the physics of star formation. 
Combining JWST mid-infrared absorption observations with {\it Herschel} submillimeter emission data makes it possible to conduct multi-scale analyses 
of the texture of molecular clouds over four orders of magnitude or more in spatial scales. This is invaluable to characterize the hierarchical nature 
of the fragmentation of molecular clouds leading to star formation and identify physically distinct levels in the hierarchy of cloud structures, 
marking changes in the dominant physics at different scales. 
We thus anticipate that further absorption studies of infrared-dark clouds with JWST/MIRI have the potential to shed new light on the intimate links 
between the structure of the cold ISM and the star formation process.

\begin{acknowledgments}
We thank the referee for constructive comments which helped us strengthen the paper.
Ph.A. and M.M. acknowledge financial support by CNES and “Ile de France” regional funding (DIM-ACAV+ Program). 
AZ thanks the support of the Institut Universitaire de France. We also acknowledge support from the French Thematic Actions 
of CNRS/INSU on stellar and ISM physics (ATPS and PCMI), co-funded by CNRS, CEA, and CNES.
We are grateful to E. Schisano for kindly providing the 
filament width measurements from \citet{Schisano+2014}, used 
in Figure~\ref{width_vs_resol} of this paper. 
This study has made use of data from the HOBYS project, a {\it Herschel} Key Program carried out by 
SPIRE Specialist Astronomy Group 3 (SAG 3), scientists of the LAM laboratory in Marseille, and scientists of the Herschel Science Center (HSC).
\end{acknowledgments}

%

\vspace{5mm}
\facilities{JWST(MIRI), APEX(ArT\'eMiS), \hbox{{\it Herschel} (SPIRE and PACS)}, {\it Spitzer}}








\pagebreak
\newpage

\bibliography{filaments_ref}{}
\bibliographystyle{aasjournal}



\appendix

\section{{Evaluation of potentially saturated areas}}
\label{app:saturation}

{To assess whether our absorption maps are potentially saturated in the highest column density areas, 
we compared both the JWST intensities (Figures~\ref{ngc6334_jwst_images}a and \ref{ngc6334_jwst_images}b) and the JWST-based column densities 
($N_{\rm H_2,\, 7.7\mu \rm{m}}^{\rm JWST2}$ and $N_{\rm H_2,\, 25.5\mu \rm{m}}^{\rm JWST2}$, cf. Figure~\ref{ngc6334_jwst_coldens_images}c)
to the column density estimates independently derived with {\it Herschel}\,$+$\,APEX at $8\arcsec $ resolution in a strip of width $\sim$\,0.1\,pc 
(or $\sim$\,$16\arcsec$) following the crest of the NGC~6334M filament. 
This was done after smoothing and resampling the JWST data 
to the same resolution and Nyquist pixel size as the APEX data.} 

\begin{figure}[!hb]
\centering
\resizebox{0.99\hsize}{!}{\includegraphics[angle=0]{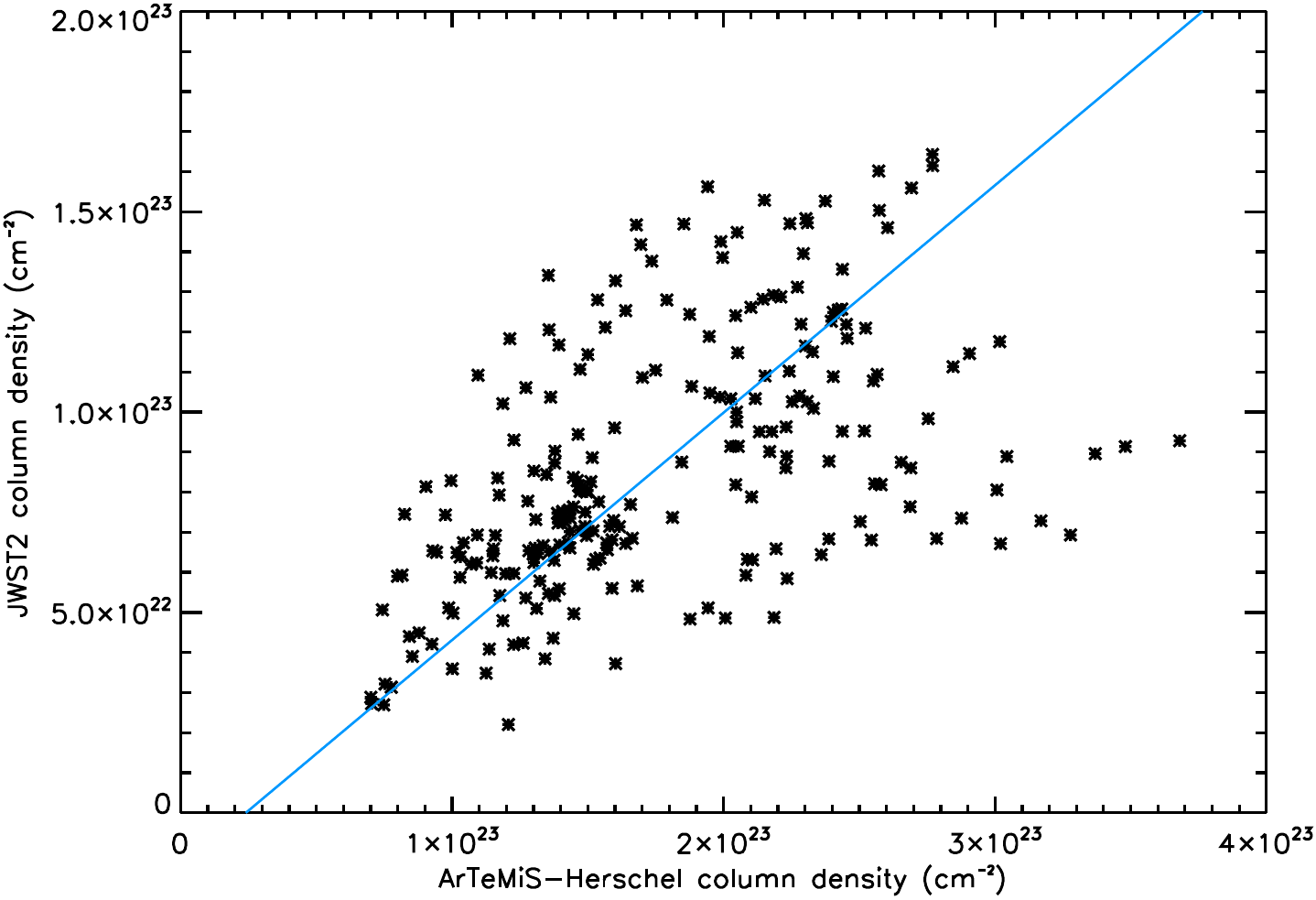}} 
\caption{{Comparison between JWST column density estimates from method 2 ($N_{\rm H_2,\, 7.7\mu \rm{m}}^{\rm JWST2}$) 
and {\it Herschel}$+$APEX estimates ($N_{\rm H_2}^{\rm ArTeMiS}$) in a 0.1-pc-wide strip along the crest of the NGC~6334M filament.
The blue line is a linear bisector (or robust) least-square fit to the data points, which has a slope of $0.6 \pm 0.3$.}
}
\label{jwst_artemis_correl}
\end{figure}

{Figure~\ref{jwst_artemis_correl}, which plots $N_{\rm H_2,\, 7.7\mu \rm{m}}^{\rm JWST2}$ against $N_{\rm H_2}^{\rm ArTeMiS}$, 
displays one of these comparisons. 
Despite some scatter, a very significant ($>$\,8$\sigma$) correlation can be seen, with a linear slope of $60\% \pm 30\%$. 
We regard this as a satisfactory correlation given that the $N_{\rm H_2}^{\rm ArTeMiS}$ map is itself uncertain by a factor of $\ga 2$ 
at the highest column densities $\ga 10^{23}\, \rm{cm}^{-2}$ (see discussion in \citealp{Schuller+2021}).
On average, the two independent column density estimates $N_{\rm H_2,\, 7.7\mu \rm{m}}^{\rm JWST2}$ and $N_{\rm H_2}^{\rm ArTeMiS}$ 
agree within a level of uncertainty comparable to the intrinsic $\sim$\,20--50\% uncertainty affecting our JWST-based column density maps (cf. Sect.~\ref{sec:obs}), 
even close to the crest of the filament.
There is some indication 
that the $N_{\rm H_2,\, 7.7\mu \rm{m}}^{\rm JWST2}$--$N_{\rm H_2}^{\rm ArTeMiS}$ correlation 
becomes weaker for the data points with $N_{\rm H_2}^{\rm ArTeMiS} \ga 2.5 \times 10^{23}\, \rm{cm}^{-2}$, 
which may result from saturation above this column density. 
Given our dust opacity assumptions, this would correspond to $\tau_{7.7\mu{\rm m}} \sim 5$, which compares well 
with the saturation optical depths estimated in similar absorption studies with {\it Spitzer} \citep[e.g.,][]{Peretto+2009}. 
Likewise, $N_{\rm H_2,\, 25.5\mu \rm{m}}^{\rm JWST2}$ is significantly correlated with $N_{\rm H_2}^{\rm ArTeMiS} $ 
along the filament crest (at the $>$\,4$\sigma$ level), with some indication of saturation effects for $N_{\rm H_2}^{\rm ArTeMiS} \ga 2 \times 10^{23}\, \rm{cm}^{-2}$. 
Conversely, as mentioned in Sect.~\ref{sec:obs}, the JWST $7.7\mu \rm{m}$ and $25.5\mu \rm{m}$ intensities 
are strongly anti-correlated with $N_{\rm H_2}^{\rm ArTeMiS} $ along the filament crest.
}

\begin{figure}
\centering
\resizebox{0.99\hsize}{!}{\includegraphics[angle=0]{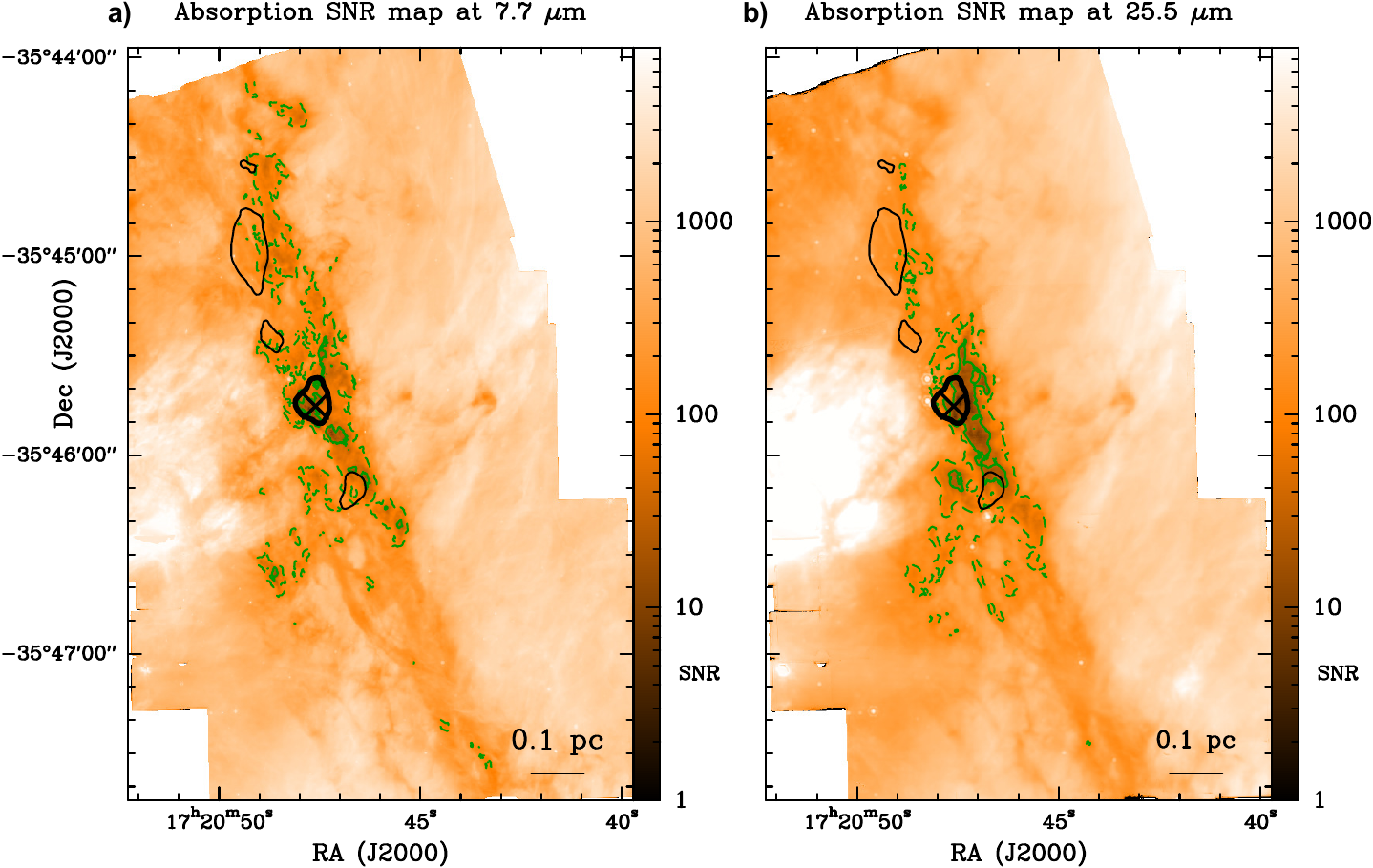}} 
\caption{{Maps showing the estimated {instrumental} signal-to-noise ratio (SNR) of the JWST absorption maps at 7.7\,$\mu$m {\bf (a)} and 25.5\,$\mu$m {\bf (b)}. 
The solid and dashed green contours mark instrumental SNR levels of 30 and 110, 
corresponding approximately to SNR levels of $\sim$\,1 and  $\sim$\,3 with respect to the estimated level of background/foreground fluctuations on small scales 
$\delta I_{\rm rms} \sim 10$\,MJy/sr, 
respectively. A black cross near field center marks the position of the minimum intensity in the JWST maps.
For comparison, black contours show the locations of local maxima/clumps in the $N_{\rm H_2}^{\rm ArTeMiS} $ map; 
the thicker contour corresponds to an ArTeMiS clump which coincides with the JWST minimum position.}
}
\label{abs_snr_maps}
\end{figure}

{To further investigate where the JWST absorption maps may be potentially saturated, we show maps of 
$\left(I_{\rm obs}(\vec{r}) - I_{\rm fore}\right)/I_{\rm rms}$ given our estimated values of  $ I_{\rm fore}$ and $ I_{\rm rms}$  at both wavelengths 
in Figure~\ref{abs_snr_maps}.
According to Equation~(\ref{equ1}), these maps visualize the signal-to-noise ratio (SNR) of the residual absorption signal 
$I_{\rm back}(\vec{r}) \times e^{-\tau_\lambda}$. 
It can be seen that this SNR is very high at both wavelengths everywhere in the mapped region, except possibly in the north-eastern part of the filament. 
Moreover, the region of highest mid-infrared optical depth, where $I_{\rm obs} \sim I_{\rm obs}^{\rm min}$ at both wavelengths (see black cross in Figure~\ref{abs_snr_maps}), 
coincides with a dense, starless submillimeter dust continuum clump detected by APEX/ArT\'eMiS (thick black contour in Figure~\ref{abs_snr_maps}) 
and very close to the SCUBA source JCMTSF~J172047.8-354550 \citep{DiFrancesco+2008}.
The southernmost ArT\'eMiS clump shown as a black contour in Figure~\ref{abs_snr_maps} is also associated with high mid-infrared optical depth.
This is not the case for the two ArT\'eMiS clumps to the north of the JWST minimum position, 
but these appear to correspond to faint protostellar sources detected {\it in emission} by MIRI at 25.5\,$\mu$m.
We conclude that our JWST absorption maps are most likely not saturated in the south-western part of the filament and 
only marginally saturated in a limited portion of the north-eastern part. 
The MIRI $7.7\,\mu \rm{m}$ band appears to be slightly better than the $25.5\,\mu \rm{m}$ band at tracing starless regions of high column density,  
but conversely the  $25.5\,\mu \rm{m}$ band is more effective at detecting weak protostellar sources in emission.
}

\section{{Autocorrelation function analysis}}
 \label{app:ACF}

The autocorrelation coefficient function (ACF) of a given image 
$I(\vec{r})$ of NGC~6334M is defined as:
\begin{equation}
{\rm ACF}_I(\vec{\Delta r}) = \frac{ {\rm cov} \left( I, {\rm T}_{\vec{\Delta r}}(I) \right) } {V(I)},
\label{equ6}
\end{equation}
where $[T_{\vec{\Delta r}}(I)](\vec{r}) = I(\vec{r}-\vec{\Delta r}) $ is a version of the original image translated 
by a vector displacement $\vec{\Delta r}$, 
$V(I) = \langle \, \delta I(\vec{r})^2 \, \rangle $,  
where \hbox{$\delta I(\vec{r}) = I(\vec{r}) - \langle I(\vec{r}) \rangle $}, is the variance of the original image, and 
\hbox{${\rm cov}\left( I ,{\rm T}_{\vec{\Delta r}}(I) \right) = \langle \, \delta I(\vec{r})\,  \delta T_{\vec{\Delta r}}(I)(\vec{r})  \, \rangle $}
is the covariance of the two images $I$ and $T_{\vec{\Delta r}}(I)$ \citep[see, e.g.,][]{Kleiner+1984}. 
In practice, ACF$_I$ is another image with the same dimensions as the original image $I$ and 
values ranging between $-1$ and $+1$. It is symmetric about its center and informs about the 
presence of significant correlations or characteristic scales in the original image as a whole. 

Applied to images of filaments, the ACF is 
potentially a powerful tool\footnote{ACF$_I$ can be easily calculated as the 
normalized inverse Fourier transform of the power spectrum of $I$. To avoid aliasing effects, the size of the input image is  
expanded by a factor of $\ge 2$ in both dimensions, with a padding value equal to the mean of the input image.} 
to characterize average filament widths. 
However, one has to take care that the ACF characterizes the entire input image, not only its behavior on small spatial scales. 
For a filament, 
the ACF thus depends on the entire radial profile of the filament, including the outer radius, 
not only the inner portion (see Sect.~\ref{sec:conv_test} and Figure~\ref{n6334_conv_test}  for an example of this effect). 

\begin{figure}[!htb]
\centering
\resizebox{0.99\hsize}{!}{\includegraphics[angle=0]{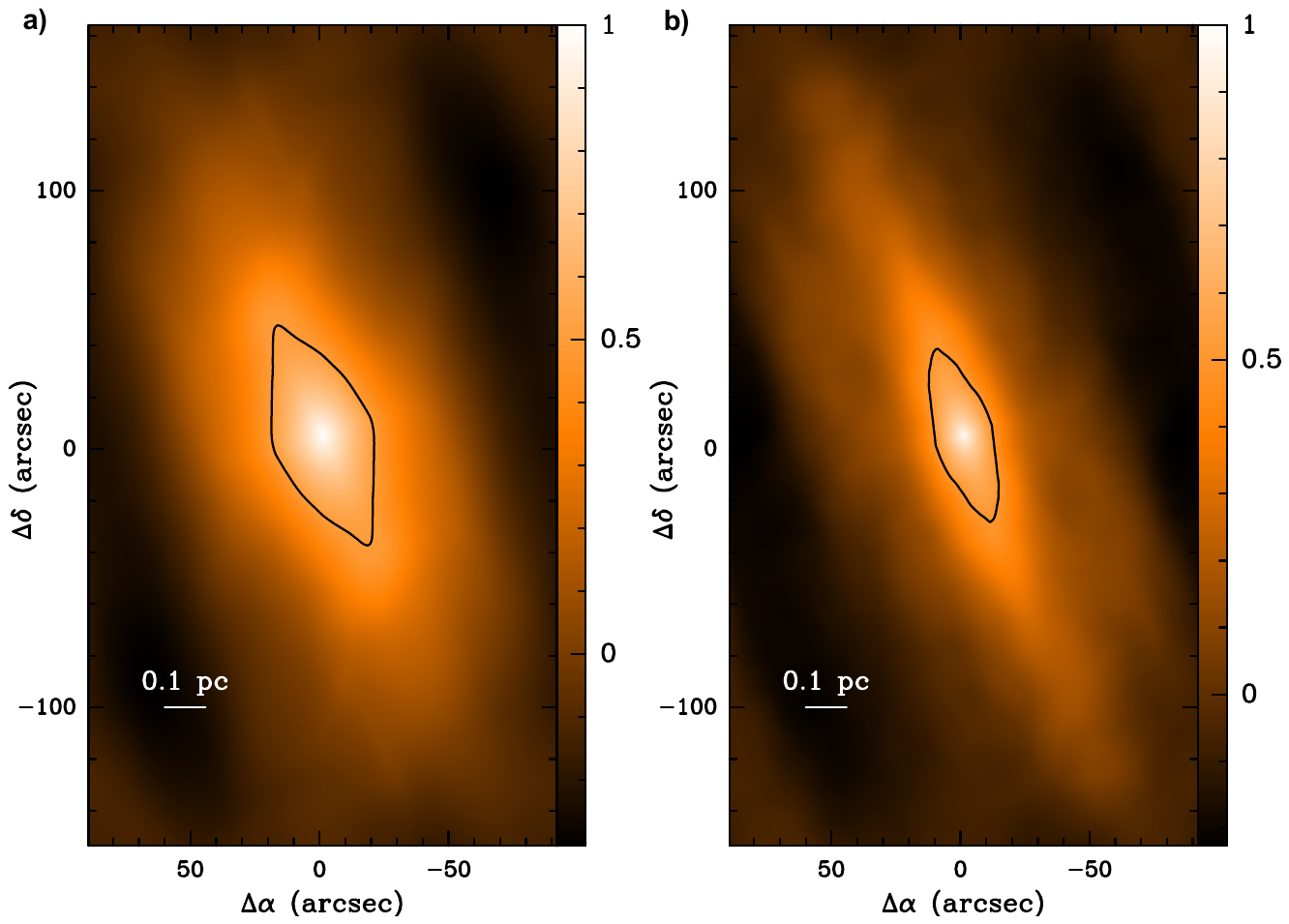}} 
\caption{Normalized autocorrelation function (ACF) maps of the JWST-based column density images 
$N_{\rm H_2,\, 7.7\mu \rm{m}}^{\rm JWST1}$ {\bf (a)} and $N_{\rm H_2,\, 7.7\mu \rm{m}}^{\rm JWST2}$ {\bf (b)}. 
[See panels d) and c) of Figure~\ref{ngc6334_jwst_coldens_images}, respectively,  for the corresponding column density images.]
In both panels, a black contour marks the 0.5 (half-maximum) level of the ACF.
}
\label{autocorrel_images}
\end{figure}

In Figure~\ref{autocorrel_images}, we show the ACF images of the two column density images 
$N_{\rm H_2}^{\rm JWST1}(\vec{r})$ and $N_{\rm H_2}^{\rm JWST2}(\vec{r})$ derived from the JWST 7.7\,$\mu$m data 
in Sect.~\ref{sec:coldens}.
As can be seen from comparison of Figure~\ref{autocorrel_images} and Figure~\ref{ngc6334_jwst_coldens_images}, 
both ACF images are elongated along the same direction as the filament. 
As estimated from the main principal axis of the second-order moment matrix of the ACF image(s), 
the mean position angle of the major axis of the NGC~6334M filament is P.A.\,$=+21\degr \pm 1\degr $ (east of north).
The {main} advantage of considering the ACF over the original image(s) is that the former is automatically ``straight'' and symmetric about its main axis, 
circumventing the need to trace the detailed path of the filament crest. 
{Moreover, the ACF 
has interesting translation invariance properties: ${\rm ACF}_{{\rm T}_{\vec{\Delta s}}(I)} = {\rm ACF}_I$ for any global displacement ${\vec{\Delta s}}$, 
and transverse ACF profiles ${\rm tACF}_I(\Delta x)$ for 
$\Delta y = 0$ are invariant under any set of lateral translations  
$\vec{\Delta s_i} = \Delta x_i\,\vec{u}_x$ operating independently on each row $i$ of the image $I$ (at $y = y_i$). 
This implies that, as long as there is a dominant filament in the image, with a relatively well-defined main principal axis (i.e., a crest deviating by less 
than, say, $\pm 45\degr$ from this axis at any position), 
a tranverse ACF cut through the minor axis of ${\rm ACF}_I$ can inform about the average filament width 
to a good approximation even if the filament is not perfectly straight.}
{Another advantage is that the ACF image has a natural base level corresponding to ${\rm ACF} = 0$ (absence of correlation), 
eliminating the need to estimate a background level when analyzing radial profiles.}

\begin{figure}[h!]
\centering
\resizebox{0.99\hsize}{!}{\includegraphics[angle=0]{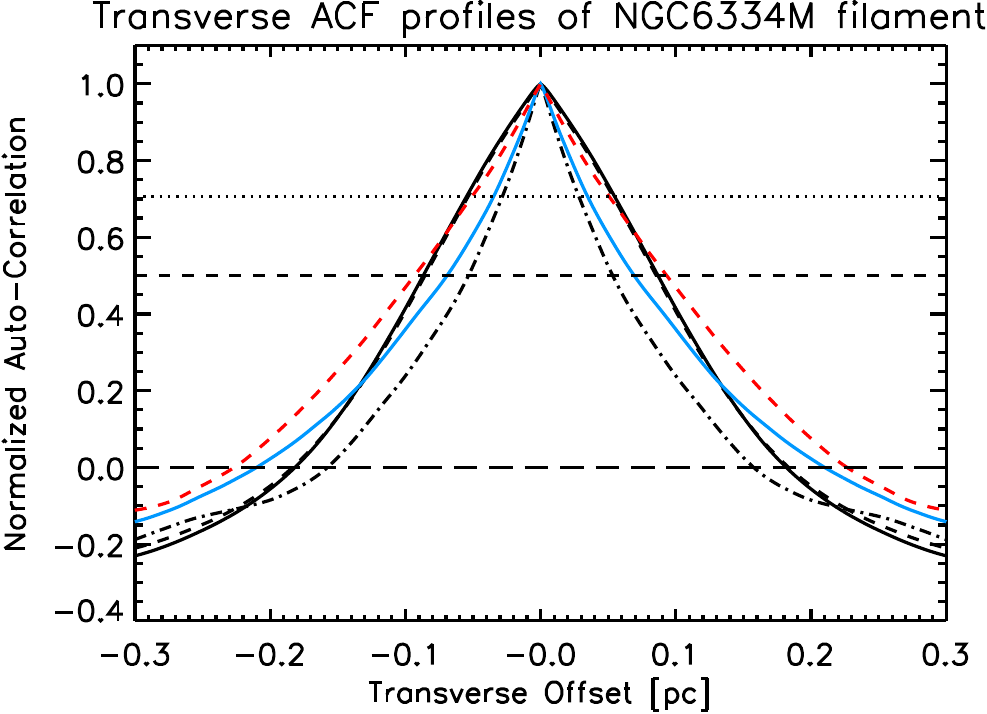}}
\caption{Transverse, autocorrelation coefficient (ACF) profiles of the NGC~6334M filament as derived 
{in the JWST mask defined in Sect.~\ref{sec:conv_test} and Figure~\ref{ngc6334_multi_resol_images}} 
from both the original JWST/MIRI 7.7\,$\mu$m (blue solid curve) and 25.5\,$\mu$m (red dashed) images 
and the column density maps $N_{\rm H_2,\, 7.7\mu \rm{m}}^{\rm JWST1}$ (black solid curve), 
$N_{\rm H_2,\, 7.7\mu \rm{m}}^{\rm JWST2}$  (black dash-dotted), and 
$N_{\rm H_2,\, 25.5\mu \rm{m}}^{\rm JWST1}$ (black dashed).
{To guide the eye, 
a long-dashed line marks the base level (${\rm ACF} = 0$), 
a short-dashed line an ACF value of 0.5 (HP level), and a dotted line an ACF level of $1/\sqrt{2}$. 
The widths of the profiles at the latter level correspond to the FWHM$_{\rm ACF}$ estimates 
of the filament width referred to in the text.}
}
\label{ACF_profiles}
\end{figure}

Figure~\ref{ACF_profiles} displays transverse cuts through the minor axis of the ACF maps 
of both the original JWST (7.7 and 25.5\,$\mu$m) images and the corresponding column density maps, 
all taken at the center of the maps (i.e., at $\Delta y = 0$, where $\Delta y $ denotes the displacement offset along the main axis of the filament). 
For a 
filament with a pure Gaussian radial profile, the HP width of 
the transverse ACF cut, $W_{\rm HP}^{\rm ACF}$, is expected to be $\sqrt{2}\times {\rm FWHM_{\rm fil}}$, 
where FWHM$_{\rm fil}$ is the HP width of the 
filament radial profile\footnote{{Strictly speaking, the ratio FWHM$_{\rm fil}/W_{\rm HP}^{\rm ACF}$ is $1/\sqrt{2}$ 
when the size of the map is large compared to the width of the Gaussian profile. 
For a Plummer profile, FWHM$_{\rm fil}/W_{\rm HP}^{\rm ACF}$ would be different and would depend on the map size. 
However, the ACF profile is generally broader than the filament profile, and $0.3 \la {\rm FWHM}_{\rm fil}/W_{\rm HP}^{\rm ACF} \la 1$ in practical cases.
{If the filament is not perfectly straight, there will also be a slight extra broadening of the transverse ACF profile compared to a straight filament.}
}}. 
In the following and in Table~\ref{table_widths}, we thus denote by FWHM$_{\rm ACF}$ the equivalent HP width of the filament 
as estimated from the transverse ACF cut assuming a Gaussian filament profile 
(FWHM$_{\rm ACF} = W_{\rm HP}^{\rm ACF}/\sqrt{2}$). 
The half-power width of the JWST ACF profile is 0.14\,pc at 7.7\,$\mu$m and 
0.19\,pc at 25.5\,$\mu$m, translating into FWHM$_{\rm ACF}$ values of 0.10\,pc and 0.13\,pc, respectively. 
The FWHM$_{\rm ACF}$ width derived from both the $N_{\rm H_2,\, 7.7\mu \rm{m}}^{\rm JWST1}$ and 
the $N_{\rm H_2,\, 25.5\mu \rm{m}}^{\rm JWST1}$ ACF profile is 0.12\,pc. 
The $N_{\rm H_2,\, 7.7\mu \rm{m}}^{\rm JWST2}$ ACF profile (dash-dotted in Figure~\ref{ACF_profiles})
is a bit narrower with a FWHM$_{\rm ACF}$ value 
of 0.08\,pc, probably due to the fact that our second method of deriving 
column density maps from absorption data (see Sect.~\ref{sec:coldens}) 
effectively filters out a fraction of the signal at large spatial scales. 
Overall, the results obtained from our ACF analysis of the JWST data agree well with 
with the filament width estimates derived 
from direct measurements of the radial 
profiles (see Figures~\ref{rad_profile} and~\ref{histo_widths}). 

{The ACF technique is also useful to characterize input images with more than one or no obvious filament. For an isotropic synthetic image 
with a scale-free power spectrum $P_{\rm 2D} (k) \propto k^{-1.6}$ similar to that observed in NGC~6334 with {\it Herschel}, for instance, 
transverse ACF measurements, made across any given axis at various resolutions, would yield results resembling those 
obtained for the pure power-law model filament (dash-dotted blue line) in Figure~\ref{n6334_conv_test} and 
would not exhibit any convergence below $\sim$\,0.1\,pc. The real NGC~6334 data are in clear contrast with this scale-free situation (see Sect.~\ref{sec:conv_test}).}

\section{Derivation of the power spectra}
\label{app:power_spectra}

The circularly-averaged power spectra shown in Figure~\ref{power_spectra} were derived using a well-proven approach 
for 2D astronomical images \citep[e.g.,][]{mamd+2010,Roy+2015}, which we briefly describe here.
For both the JWST ($N_{\rm H_2}^{\rm JWST1}$) 
and the {\it Herschel} column density map, we first constructed an observed power spectrum image by evaluating 
the Fourier transform of the input 2D data, $N_{\rm obs}(\vec{r})  $: \hbox{$P_{\rm obs}(\vec{k}) = \left| \hat{N}_{\rm obs}(\vec{k}) \right| ^2 $}, where 
$\hat{N}(\vec{k})$ denotes the 2D Fourier transform of column density map $N(\vec{r})  $. 
We then averaged over all directions to obtain an observed power spectrum profile, $P_{\rm obs}(k) $. 

Each observed map 
can be written as the sum of two independent, uncorrelated terms, \hbox{$N_{\rm obs}(\vec{r}) =  N_{\rm true}(\vec{r}) \ast B(\vec{r}) + W(\vec{r})$}, 
where $N_{\rm true}(\vec{r})$ is the intrinsic sky map, 
$B(\vec{r}) $ the telescope beam, $\ast$ denotes convolution, and $ W(\vec{r})$ is a (white) noise term.
The observed power spectrum profile  
can thus be modeled as \hbox{$P_{\rm obs}(k) = P_{\rm true}(k) \times \gamma_{\rm beam}(k) + P_W(k) $}, 
where $\gamma_{beam}(k)$ represents the power spectrum of the telescope beam and  $P_W(k) $ is the power spectrum of the noise term. 
If the beam $B(\vec{r}) $ is approximately Gaussian with a full width at half maximum FWHM$_{\rm b}$, 
then $\gamma_{\rm beam}(k)$ is also approximately Gaussian in $k$-space and its FWHM 
is FWHM$_\gamma = \sqrt{8}\ {\rm ln}\,2/(\pi\,{\rm FWHM_b}) $. 
The amplitude $\bar{P}_W$ of the white noise contribution $P_W(k) $ can be estimated as the median of $P_{\rm obs}(k) $ 
at high spatial frequencies, in the range \hbox{$0.75\, k_{\rm max} < k < k_{\rm max}$}, where $k_{\rm max} $ is the 
Nyquist spatial frequency \hbox{$k_{\rm max} = 1..22/{\rm FWHM_b} \approx 2\,{\rm FWHM_\gamma}$}. 

The power spectra shown in Figure~\ref{power_spectra} correspond to beam-corrected, noise-subtracted 
estimates of $P_{\rm true}(k)$ for both {\it Herschel} and JWST, i.e., 
\hbox{$ P_{\rm true}(k) = \left[P_{\rm obs}(k) - \bar{P}_W\right]/\gamma_{\rm beam}(k)$}. 
In both cases, $ P_{\rm true}(k)$ is well fit by a power law for spatial frequencies in the range
\hbox{$k_{\rm min} \la k \la  {\rm FWHM}_\gamma /2 $}, where \hbox{$k_{\rm min} = \frac{1}{\rm Map\, Size}$} 
is the lowest available spatial frequency, with ${\rm Map\, Size} \sim 3\arcmin$ for the JWST data and 
and ${\rm Map\, Size} \sim 75\arcmin $ for the {\it Herschel} data. 
In the range ${\rm FWHM}_\gamma /2 < k <  k_{\rm max} $, $ P_{\rm true}(k)$ becomes more uncertain 
due the strong influence of the beam and noise corrections. 
In practice, the JWST 7.7\,$\mu$m data provide reliable power spectrum information for \hbox{$0.3 \la k \la 80\, {\rm arcmin}^{-1}$}, 
while the {\it Herschel} data of \citet{Russeil+2013} constrain the range \hbox{$10^{-2} \la k \la 1$--$2\, {\rm arcmin}^{-1}$}. 
Since the {\it Herschel} column density map covers a wider area than, and does not have the same variance as, 
the JWST image, the {\it Herschel} power spectrum was scaled by the variance ratio $V_{\rm JWST}/V_{\it Herschel}$ 
for easier comparison with the JWST power spectrum in Figure~\ref{power_spectra}. 
Given this scaling, the two power spectra agree well with one another in the overlap range 
of spatial frequencies $0.3 \la k \la 1$--$2\, {\rm arcmin}^{-1}$. 

\end{document}